\documentclass[%
reprint,
floatfix,
superscriptaddress,
amsmath,amssymb,
aps,
pra,
]{revtex4-2}
\usepackage[utf8]{inputenc}
\usepackage[english]{babel}
\usepackage{blindtext}
\usepackage{amsmath}
\usepackage{xcolor,colortbl}
\definecolor{woNbPad}{RGB}{240,142,80}
\definecolor{wNbPad}{RGB}{199,0,199}
\usepackage{array}
\newcommand\Tstrut{\rule{0pt}{2.15ex}}         % = `top' strut
\newcommand\Bstrut{\rule[-0.45ex]{0pt}{0pt}}   % = `bottom' strut
\newcommand\Tstruttop{\rule{0pt}{2.5ex}}         % = `top' strut
\newcommand\Bstruttop{\rule[-0.6ex]{0pt}{0pt}}   % = `bottom' strut
\newcommand\corrs[1]{{\color{black}#1}} %%%%%%%%%%%%%%%%%%%%%%
\newcolumntype{C}{>{$\displaystyle}c<{$}}
\usepackage{graphicx}   %include figure files
\graphicspath{{figs/}}   %add figure path
\usepackage{dcolumn}% Align table columns on decimal point
\usepackage{bm}% bold math
\usepackage[hidelinks]{hyperref}% add hypertext capabilities
\usepackage{epstopdf}
\usepackage{braket}
\usepackage{textcomp}
\usepackage{upgreek}
\usepackage{amsmath}
\usepackage{siunitx}
\usepackage{changes}
\usepackage{natbib}
\usepackage[title]{appendix}
\usepackage[all]{hypcap}% Links point to start of figures instead of captions
\usepackage[capitalise,nameinlink]{cleveref}

\usepackage{multirow}

\begin{document}
\title{Granular~Aluminum~Parametric~Amplifier~for~Low-Noise~Measurements~in~Tesla~Fields}

\author{Nicolas~Zapata}
\email{nicolas.gonzalez2@kit.edu}
\affiliation{IQMT,~Karlsruhe~Institute~of~Technology,~76131~Karslruhe,~Germany}

\author{Ivan~Takmakov}
\affiliation{IQMT,~Karlsruhe~Institute~of~Technology,~76131~Karslruhe,~Germany}
\affiliation{PHI,~Karlsruhe~Institute~of~Technology,~76131~Karlsruhe,~Germany}
\affiliation{Current address:~IQM~Quantum~Computers,~Espoo~02150,~Finland}

\author{Simon~Günzler}
\affiliation{IQMT,~Karlsruhe~Institute~of~Technology,~76131~Karslruhe,~Germany}
\affiliation{PHI,~Karlsruhe~Institute~of~Technology,~76131~Karlsruhe,~Germany}

\author{Simon~Geisert}
\affiliation{IQMT,~Karlsruhe~Institute~of~Technology,~76131~Karslruhe,~Germany}

\author{Soeren~Ihssen}
\affiliation{IQMT,~Karlsruhe~Institute~of~Technology,~76131~Karslruhe,~Germany}

\author{Mitchell~Field}
\affiliation{IQMT,~Karlsruhe~Institute~of~Technology,~76131~Karslruhe,~Germany}

\author{Ameya~Nambisan}
\affiliation{IQMT,~Karlsruhe~Institute~of~Technology,~76131~Karslruhe,~Germany}

\author{Dennis~Rieger}
\affiliation{PHI,~Karlsruhe~Institute~of~Technology,~76131~Karlsruhe,~Germany}

\author{Thomas~Reisinger}
\affiliation{IQMT,~Karlsruhe~Institute~of~Technology,~76131~Karslruhe,~Germany}

\author{Wolfgang~Wernsdorfer}
\affiliation{IQMT,~Karlsruhe~Institute~of~Technology,~76131~Karslruhe,~Germany}
\affiliation{PHI,~Karlsruhe~Institute~of~Technology,~76131~Karlsruhe,~Germany}

\author{Ioan~M.~Pop}
\email{ioan.pop@kit.edu}
\affiliation{IQMT,~Karlsruhe~Institute~of~Technology,~76131~Karslruhe,~Germany}
\affiliation{PHI,~Karlsruhe~Institute~of~Technology,~76131~Karlsruhe,~Germany}
\affiliation{Physics~Institute~1,~Stuttgart~University,~70569~Stuttgart,~Germany}

\date{\today}
%-----------------------------------------------------------------------------

\begin{abstract}
Josephson junction parametric amplifiers have become essential tools for microwave quantum circuit readout with minimal added noise. Even after improving at an impressive rate in the last decade, they remain vulnerable to magnetic field, which limits their use in many applications such as spin qubits, Andreev and molecular magnet devices, dark matter searches, etc. Kinetic inductance materials, such as granular aluminum (grAl), offer an alternative source of non-linearity with innate magnetic field resilience. We present a non-degenerate amplifier made of two coupled grAl resonators resilient to in-plane magnetic field up to \SI{1}{T}. 
It offers 20~dB of gain close to the quantum limit of added noise, with a gain-bandwidth product of 28~MHz and -110~dBm input saturation power.
\end{abstract}

%-----------------------------------------------------------------------------

\maketitle

%-----------------------------------------------------------------------------
The remarkable progress in Josephson junction (JJ) based parametric amplifiers sparked their adoption in both academic and industrial research groups, establishing single-shot quantum state detection of superconducting qubits as an integral capability in circuit quantum electrodynamics \cite{Vijay2011Mar,Roy2016Aug,Aumentado2020Jul, Esposito2021Sep}. State-of-the-art amplifiers are essential for syndrome measurements during error correction of superconducting quantum registers \cite{Krinner2022May,White2023Jan,BibEntry2023Feb,Sivak2023Apr}, they enable the development of novel detection techniques \cite{Vesterinen2017Jun,Minev2019Jun,Kokkoniemi2019Oct,Wang2023Jul,Assouly2023Oct}, and they serve as indispensable instruments to uncover microscopic mechanisms degrading superconducting qubit coherence \cite{Murch2013Oct,Serniak2019Jul,Spiecker2023Sep}. 
However, the magnetic field induced Fraunhofer pattern for the JJ critical current \cite{Tinkham, Schneider2019Sep,Krause2022Mar} constrains their applicability in systems such as spin \cite{Wang2023Jul,Schaal2020Feb,Elhomsy2023Jul} and Andreev qubits \cite{Pita-Vidal2023Aug,Hays2021Jul}, spin ensembles \cite{Bienfait2016Mar,Vine2023Mar}, molecular magnet devices \cite{Godfrin2017Nov} and dark matter detectors \cite{Lamoreaux2013Aug,Backes2021Feb,Braggio2022Sep}.

Recent experiments have used magnetic shielding to protect JJ-based amplifiers~\cite{Elhomsy2023Jul,Wang2023Jul,Janssen2024Feb}, at the price of placing the amplifier farther from the measured device and risking to reduce the readout efficiency. Ideally, a magnetic field resilient amplifier is directly coupled to the device under test. Graphene-based amplifiers~\cite{Sarkar2022Nov,Butseraen2022Nov} and devices employing proximitized semiconductors~\cite{Phan2023Jun, Splitthoff2023Aug} have been developed to achieve this goal. However, technologically involved fabrication makes their integration into more complex systems challenging and their relatively low saturation power (-130 to -120~dBm) limits the readout signal amplitude. An alternative path employs the non-linearity of kinetic inductance materials like NbN~\cite{Xu2023Feb,Frasca2023Dec}, NbTiN~\cite{Khalifa2023Mar,Parker2022Mar,Vaartjes2023Nov, Mohamed2023Nov} or granular aluminum (grAl)~\cite{Maleeva2018,Borisov2020}, which offer magnetic field resilience, adaptable integration and larger dynamic range. 

Remarkable progress has recently been reported for NbN amplifiers, which were proven to operate close to the quantum limit up to 0.5~T~\cite{Xu2023Feb}. Even though NbN enabled these achievements, its relatively large critical current density, in the range of~mA/\SI{}{\micro \metre}$\mathrm{^2}$, requires pushing the boundaries of device miniaturization to reach non-linearity values comparable to JJ devices. Two orders of magnitude lower critical current densities and higher non-linearity can be achieved in grAl~\cite{Maleeva2018,Borisov2020, Winkel2020Transmon}, while remaining an order of magnitude below the superconducting-to-insulator transition (SIT) threshold~\cite{Bachar2013Jun,Pracht2017Sep,Levy-Bertrand2019Mar,Yang2020Sep}.

In this work we tailor the non-linearity of grAl to fabricate a standing wave parametric amplifier (grAlPA), which operates near the quantum limit in magnetic fields up to 1~T in-plane. We obtain 20~dB amplification, 28~MHz gain-bandwidth product and a saturation power of -110~dBm, satisfying typical requirements for quantum device readout. These features make grAlPAs convenient and powerful tools, ready to be integrated and to extend the range of quantum limited readout applications.
 
\begin{figure*}[!t]
\includegraphics[width=6.67in]{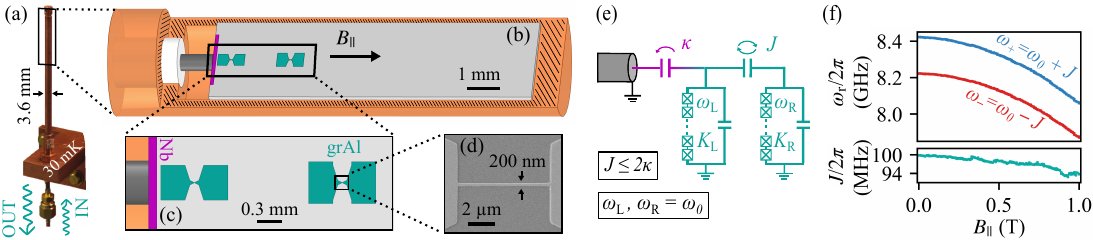}
\caption{\textbf{Device design and frequency dependence in magnetic field up to 1 T.} \textbf{(a)-(b)} Photograph and schematics of the sample holder used for microwave reflection measurements, similar to Refs. \cite{Borisov2020, Rieger2023, Rieger2023fano}. \textbf{(c)}~The granular Aluminum Parametric Amplifier (grAlPA) consists of a pair of lumped-element grAl resonators (green), comprised of two coplanar capacitor plates connected by a central strip. To stabilize the evanescent field coupling of the grAlPA to the coaxial central pin (black) we add a Nb pad (purple) at the edge of the substrate. \textbf{(d)}~Scanning electron beam image of the central grAl nanostrip. \textbf{(e)}~Equivalent circuit diagram of the grAlPA. The design resembles a Bose-Hubbard dimer, similar to Refs. \cite{Eichler2014Dimer, Winkel2020DJA}. The grAl strip shown in (c) can be modeled as an array of Josephson Junctions \cite{IvanThesis2022, Maleeva2018} with a self-Kerr non-linearity denoted as $K_i$~(i~=~R,~L). The hopping interaction $J$ gives rise to a pair of hybridized modes  $\omega \mathrm{_{\pm}}$. \textbf{(f)}~Measured frequency dependence of the dimer $\omega \mathrm{_{\pm}}$ (top) and hopping interaction $J$ (bottom) as a function of in-plane magnetic field up to 1 T.}
\label{fig_design}
\end{figure*}

The grAlPA amplifier, operated in reflection, consists of two capacitively coupled grAl resonators mounted in a single-port cylindrical waveguide, coupled via evanescent electric field to a coaxial cable (cf.~\cref{fig_design}(a)-(b)). The resonators are formed by two coplanar capacitor plates connected by a central grAl strip (cf. \cref{fig_design}(c)-(d)) which gives 80$\%$ of the total inductance~(see supplementary material). The dimensions of the capacitor plates are chosen to match the frequencies of the two modes and target perfect hybridization. The device is fabricated in a single step electron beam lithography, followed by zero-angle evaporation on a c-plane sapphire substrate~(see supplementary material). We deposit an \SI{830}{\micro \ohm}$\,$cm resistivity grAl film, 40~nm thick, which gives 120~pH/$\square$ of kinetic inductance.

The waveguide provides a clean microwave environment for the grAlPA thanks to its $\approx$~60~GHz cut-off frequency, a factor five above its operational range. The challenge when using this strategy arises from the exponential susceptibility of the coupling to variations in the distance between the pin and the grAlPA. In order to mitigate this effect we add a 40~\SI{}{\micro \meter} wide and 40~nm thick Nb pad at the edge of the chip using optical lithography. The Nb coupling pad acts as an on-chip extension of the coaxial cable central pin, uniformly spreading the electric field lines towards the grAlPA~(see supplementary material). 

Figure \ref{fig_design}(e) shows the equivalent grAlPA circuit, which can be modeled by the following Bose-Hubbard-dimer Hamiltonian \cite{Sarchi2008, Eichler2014Dimer, Winkel2020DJA}: 

\begin{equation}
H/\hbar=\sum_{i= \mathrm{L,R}}\left(\omega_ia^{\dagger}_{i}a_{i} + \frac{K_i}{2}a^{\dagger}_{i}a^{\dagger}_{i}a_{i}a_{i}\right)+Ja^{\dagger}_{\mathrm{L}}a_{\mathrm{R}} + \mathrm{h.c.},
\label{eq_full_H}
\end{equation}
where the two bosonic modes $a_{i}$, with $i$ = L, R, correspond to the left and right grAl resonators, respectively. The individual resonant frequencies are $\omega_i$, the on-site potentials $K_i$ are given by the first-order Kerr non-linearity of the central grAl strips, and the hopping term $J$ arises from the electric dipole-dipole interaction between the resonators. In case of perfect hybridization $\omega \mathrm{_L}= \omega \mathrm{_R} = \omega \mathrm{_0}$ , this interaction leads to the formation of the hybridized dimer modes $\omega \mathrm{_\pm}$ = $\omega \mathrm{_0} \pm J$, each sharing the total coupling $\kappa$ of the left resonator to the microwave environment~(see supplementary material). When a pump is applied between the hybridized modes, 4-wave mixing converts two pump photons into one signal and one idler photon in the two dimer modes, respectively, implementing a phase-preserving amplifier with signal-pump detuning~$J$.

While \cref{eq_full_H} has the necessary ingredients for parametric amplification, in order to achieve a practically useful device, two conditions need to be fulfilled. Firstly, the Kerr coefficients $K_i/2\pi$ should be $\approx$~1-100~kHz, as higher values reduce the amplifier compression point below the power commonly needed for device readout ($\approx$~-140~dBm) \cite{Eichler2014Dec,Sivak2020} and sub-kHz coefficients require higher pump powers for 20~dB gain performance,
which increase the heat load on the cryostat \cite{Hougland2024Feb}. By using a grAl strip \SI{0.2}{\micro \meter} wide and \SI{7}{\micro \meter} long, we obtain $K_i/2\pi$~=~2$\pm$1~kHz~(see supplementary material). Secondly, for both modes to be efficiently driven by the same pump, the coupling strength should satisfy the condition $J \lesssim 2\kappa$~\cite{IvanThesis2022}. At the same time, to maximize the gain-bandwidth product, we aim to increase $\kappa$ as much as possible. We design the grAlPA to have $J/2\pi$~=~0.1~GHz and $\kappa/2\pi$~=~60~MHz. 

When exposed to an in-plane magnetic field, the dimer shifts down in frequency as shown in the upper panel of \cref{fig_design}(f), tuning by approximately 300 MHz at $B_\mathrm{||}$~=~1~T, without suffering from added losses~(see supplementary material). This shift is explained by an increase of the grAl kinetic inductance, corresponding to a suppression of the superconducting gap (cf. Refs.~\cite{Borisov2020, Rieger2023, Winkel2020Transmon}~and~supplementary material). We mechanically align the field to within 1\% in-plane and we compensate in-situ the remaining out-of-plane component using a custom-designed 2D vector magnet~(see supplementary material). Importantly, the hopping strength $J$ changes by less than 10\% over the entire field range, preserving the conditions for parametric amplification up to 1~T~(cf. lower panel of \cref{fig_design}(f)).

\begin{figure}[!t]
\includegraphics[width = 1\columnwidth]{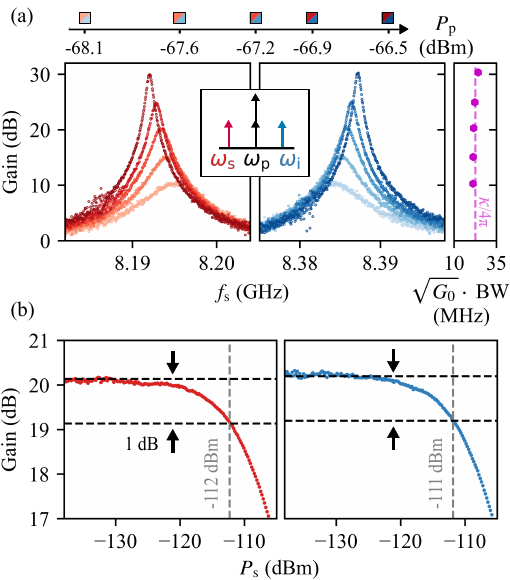}
\caption{\textbf{Gain performance in zero field.} \textbf{(a)}~GrAlPA gain vs signal frequency $f\mathrm{_s}$, for different pump powers $P\mathrm{_p}$ and fixed pump frequency $f\mathrm{_p}$ = 8.29 GHz. The inset shows a schematic of the four-wave mixing process: two pump photons (black) are converted into a signal photon (red) and an idler photon (blue), each populating a dimer mode. The right hand panel shows the measured gain-bandwidth product. \textbf{(b)}~Saturation power measurements for the grAlPA modes. $P\mathrm{_s}$ is the input signal power and the grey labels indicate the corresponding to 1-dB compression point.}
\label{fig_gain_performance}
\end{figure}

Figure~\ref{fig_gain_performance}(a) shows the gain performance of the grAlPA in zero magnetic field, as a function of increasing pump power at fixed frequency. We define the gain of the amplifier as the ratio between the reflected power with the pump on and off. Note that by this choice possible, internal losses produce asymmetric gain profiles. By increasing the pump power we can obtain a maximum gain $G_\mathrm{0}$ of 30~dB before entering the multistable regime~(see supplementary material). We measure a constant gain-bandwidth product $\sqrt{G_\mathrm{0}}\cdot \mathrm{BW} = (\kappa/2)/2\pi = $ 22.5~MHz, as illustrated in the right hand panel of \cref{fig_gain_performance}(a). Conveniently, the signal-pump detuning is in the 0.1~GHz range, as expected from the design value for the hopping term $J$, facilitating pump tone filtering at the output of the grAlPA.

The dynamic range of the grAlPA is quantified by the input signal-probe power for which the maximum gain $G\mathrm{_0}$ decreases by 1~dB, i.e. the 1-dB compression point. For this purpose, we calibrate the attenuation of the input line using the resonance fluorescence of a frequency-tunable transmon qubit \cite{Honigl-Decrinis2020Feb,GunzlerMagnetometer} in a separate experiment, employing the same setup and sample holder utilized for the grAlPA. We sample several frequencies in the interval of interest and interpolate in between, thereby calibrating the line attenuation in a 1.5~GHz frequency range~(see supplementary material). As presented in \cref{fig_gain_performance}(b), for $G_\mathrm{0}$~=~20~dB, we obtain 1-dB compression points of -111$\pm$2~dBm in both modes. This dynamic range is on par with previous implementations using arrays of Josephson junctions \cite{Winkel2020DJA, Frattini2017,Frattini2018Nov,Sivak2020} and can be improved by increasing the coupling $\kappa$ to match state-of-the-art devices~\cite{White2023Jan,Kaufman2023May}.

\begin{figure*}[t!]
\includegraphics[width=6.4in]{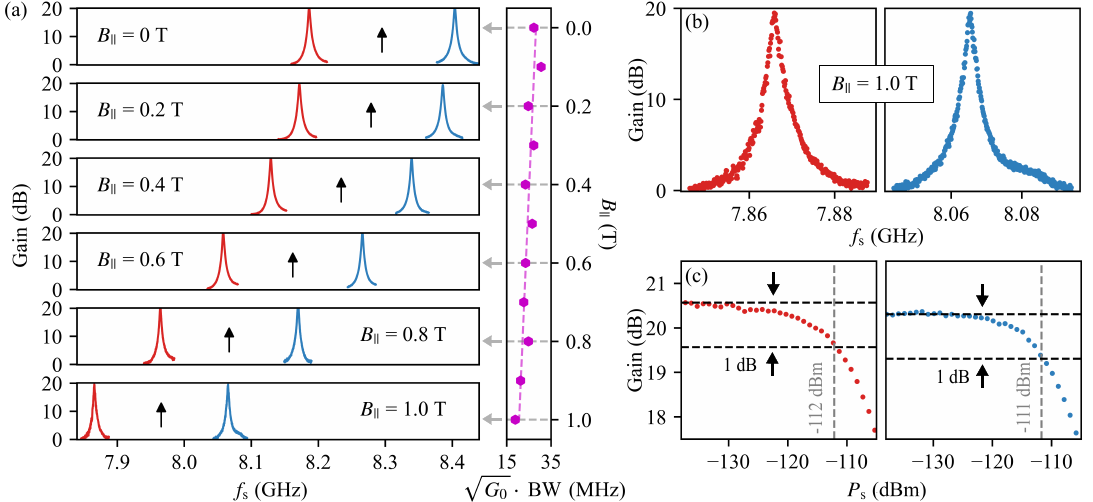}
\caption{\textbf{Gain performance resilient to in-plane magnetic field.} \textbf{(a)}~The six left panels show measured gain profiles at magnetic fields $B_\mathrm{||}$ = 0, 0.2, 0.4, 0.8, and 1.0~T, applied in parallel to the grAl film. The black arrows indicate the frequency of the pump. Note that $G_\mathrm{0}$~=~20 dB can be achieved at any field up to 1~T. The right panel depicts the gain-bandwidth product vs $B_\mathrm{||}$, extracted from the measured gain profiles with $G_\mathrm{0}$~=~20~dB. The dashed purple line illustrates the $\kappa$/2 values measured from the pump off resonant response of the dimer frequencies. \textbf{(b)}~Measured gain of the grAlPA at $B_\mathrm{||}$ = 1~T. \textbf{(c)}~Saturation power measurements for the two modes of the grAlPA at $B_\mathrm{||}$ = 1~T. The 1-dB compression point is indicated by the grey labels.}
\label{fig_gain_field}
\end{figure*}

In \cref{fig_gain_field}(a) we demonstrate that the grAlPA can be successfully operated at $G_\mathrm{0}$~=~20~dB gain in magnetic fields up to 1~T. Due to the field dependence of the dimer frequency, at each field bias the frequency of the pump needs to be adjusted, as indicated by the arrows in \cref{fig_gain_field}(a). The measured 30$\%$ decrease of $\sqrt{G\mathrm{_0}}\cdot$BW from zero field to 1~T shown in \cref{fig_gain_field}(a) is explained by the progressive decoupling of the dimer as the modes move away from the cylindrical waveguide cut-off~(see supplementary material). In \cref{fig_gain_field}(b) and (c) we present measured 20~dB gain profiles in $B_\mathrm{||}$ = 1~T and their corresponding power saturation. We find it remarkable that the grAlPA performance at $B_\mathrm{||}$ = 1~T is comparable to that at $B_\mathrm{||}$ = 0~T. The shoulder visible in the lineshape of the high frequency mode is explained by a 1~dB dip in the background signal measured without the pump, due to internal losses in the grAl resonators~(see supplementary material).

\begin{figure*}[t!]
\includegraphics[width=6.67in]{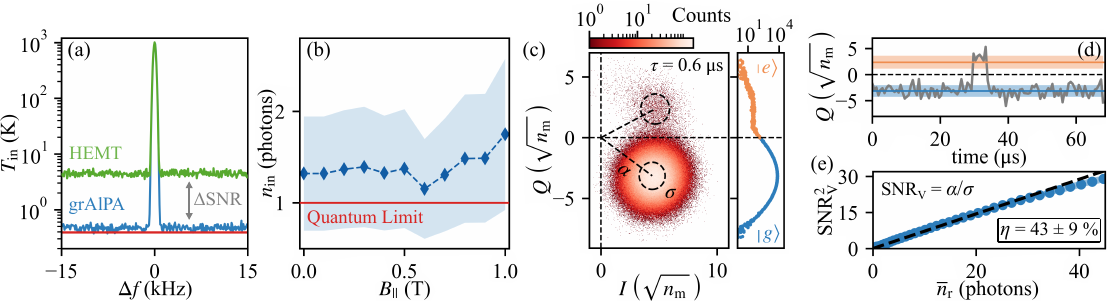}
\caption{\textbf{Noise performance and quantum efficiency of the grAlPA.} \textbf{(a)}~Input-referred noise temperature at~$B_\mathrm{||}$~=~0~T as a function of detuning $\Delta f$ from a power-calibrated tone at $f\mathrm{_s}$~=~8.193~GHz. The green and blue solid lines represent measurements with the grAlPA pump off and on, respectively. When the grAlPA is on, the noise approaches the standard quantum limit for non-degenerate amplification indicated by the solid red line. The signal-to-noise improvement ($\Delta$SNR) is defined as the ratio between the baselines of the green and blue curves. \textbf{(b)}~Input-referred noise as a function of $B_\mathrm{||}$. The blue shaded area indicates the uncertainty propagated from the line-attenuation calibration. \textbf{(c)}~Histogram of contiguous measurements of a qubit acquired using the grAlPA with $G\mathrm{_0}$~=~25~dB. The I and Q quadratures of the readout signal are scaled in units of square root of measurements photons $\sqrt{n\mathrm{_m}}$ = $\sqrt{\overline{n}_\mathrm{_r}\tau\kappa\mathrm{_r}/4}$. The Q-axis projected histogram on the right shows measured qubit distributions corresponding to the ground $|g\rangle$ and excited $|e\rangle$ state. \textbf{(d)}~Example of a measured quantum-jump trace. The solid lines correspond to the ground (blue) and excited (orange) states of the qubit and the colored areas represent one standard deviation $\sigma$ from the histogram in (e). \textbf{(e)} SNR$\mathrm{^2_{V}}$ as a function of readout resonator occupation $\overline{n}_\mathrm{_r}$. From a linear fit (black dashed line) we extract the readout quantum efficiency $\eta$~=~43~$\pm$~9~$\%$.}
\label{fig_noise_field}
\end{figure*}

Arguably the most important figure of merit of parametric amplifiers is the improvement they provide to the signal-to-noise ratio (SNR) of a readout chain. In Fig~\ref{fig_noise_field}(a), we illustrate the near-quantum limited noise performance of the grAlPA at 0~T. We measure the power spectral density (PSD) of the output line connected to the grAlPA, with the pump on and off. We convert the PSD to temperature scale, using a power-calibrated pilot tone at $f\mathrm{_s}$~=~8.193~GHz~(see supplementary material). When the pump is off, the $\approx$~4~K noise floor is set by a High Electron Mobility Transistor (HEMT). By operating the grAlPA at 20~dB gain we obtain 10~dB of SNR improvement ($\Delta$SNR), corresponding to an input-referred noise close to the standard quantum limit for phase preserving amplifiers~\cite{Caves1982Oct}. 

The grAlPA noise performance under in-plane magnetic field up to $B_\mathrm{||}$~=~1~T is shown in \cref{fig_noise_field}(b). For each point we fix the gain of the amplifier to $G\mathrm{_{0}}$~=~20~dB and tune the power-calibrated pilot tone to match the shift of the gain profiles in field (cf.~\cref{fig_gain_field}(a)). We report the input referred noise $n\mathrm{_{in}}$ in units of photons, calculated using the relation $n\mathrm{_{in}}=k\mathrm{_B}\overline{T}\mathrm{_{in}}$/$hf\mathrm{_s}$, where $\overline{T}\mathrm{_{in}}$ is the noise temperature floor extracted from the PSD with the grAlPA on. The quantum limit is given by $n\mathrm{_{in}}$~=~1, where half a photon is added by the amplifier and half a photon originates from input vacuum fluctuations. In zero field we obtain $n\mathrm{_{in}}$~=~1.21~$\pm$~0.63, corresponding to an effective grAlPA added noise $n\mathrm{_{gr}}$~=~0.71~$\pm$~0.63. Remarkably, the grAlPA remains near quantum-limited up to $B_\mathrm{||}$~=~1~T, illustrating its utility for quantum readout in magnetic field. We attribute the upturn visible in the grAlPA noise above 0.6~T to spurious vortices in the capacitor pads, which can create additional loss channels \cite{Bothner2011, Kroll2019} and should be suppressed in future designs. 

We demonstrate the utility of the grAlPA in the readout of quantum devices by showing single-shot heterodyne detection of a Generalized Flux Qubit (GFQ). The GFQ is inductively coupled to a readout resonator with linewidth $\kappa\mathrm{_r}/2\pi$~=~1.25~MHz and frequency $f\mathrm{_r}$~=~8.1361~GHz (for more details see Ref. \cite{Geisert2024Jul} device q8). In \cref{fig_noise_field}(c) we present an histogram of \num{8e5} contiguous measurements using a readout pulse of time $\tau$~=~0.6~\SI{}{\micro \second} and power equivalent to $\overline{n}_\mathrm{_r}$~=~31 circulating photons in the resonator, calibrated using the AC stark shift of the GFQ~(see supplementary material). The distribution shows two maxima corresponding to the ground $|g\rangle$ and first excited $|e\rangle$ energy-states of the qubit. Using the grAlPA with $G\mathrm{_0}$~=~25~dB, in \cref{fig_noise_field}(d) we plot a typical measured quantum-jumps trace corresponding to an effective qubit temperature $T\mathrm{_q}$~=~56~mK. We define the quadrature SNR$\mathrm{_V}$ as the ratio between the magnitude $\alpha$ and standard deviation $\sigma$ of the $|g\rangle$ pointer state. We fit the data using the formula SNR$\mathrm{^2_V}$~=~$\eta \overline{n}_\mathrm{_r} \kappa\mathrm{_r} B\mathrm{^{-1}}/4$, where $B\mathrm{^{-1}}$~=~$2/\kappa\mathrm{_r}$~+~$\tau$ is the measurement bandwidth \cite{Vijay2011Mar,Takmakov2021Jun}, from which we extract a \corrs{quantum efficiency $\eta$~=~43~$\pm$~9$\%$} (cf. \cref{fig_noise_field}(e)). Considering the contributions of the insertion loss after the GFQ and the residual HEMT noise, we extract an \corrs{intrinsic grAlPA quantum efficiency $\eta\mathrm{_{gr}}$~=~92~$\mathrm{^{+8}_{-22}} \%$, corresponding to $n\mathrm{_{gr}}$~=~0.54~$\pm$~0.13 added noise photons~(see supplementary material).}

In summary, we have demonstrated a practically useful parametric amplifier, nicknamed grAlPA, which relies exclusively on granular aluminum for its nonlinearity and operates close to the quantum limit up to one Tesla in-plane magnetic field. Beyond this range, spurious frequency jumps and added losses, likely resulting from residual flux trapping in the pads of the capacitors, hinder the operation of the amplifier. A potential solution is the implementation of artificial pinning sites, similar to the approach described in Refs. \cite{Bothner2011, Kroll2019}. Furthermore, in order to enable enhanced frequency tunability and increased signal-pump detuning, the grAlPA concept could be expanded to 3-wave mixing by using RF-squids \cite{Sivak2020, Frattini2017, Sivak2019, Miano2022,Naaman2017} made of grAl nanojunctions \cite{Rieger2023} or by applying a direct current bias \cite{Frasca2023Dec}.

The data supporting the findings presented in this Letter
are openly available \cite{zenodo_link}. 

\begin{acknowledgments}
We are grateful to Nicolas Roch, Patrick Winkel and Mathieu Fechant for fruitful discussions and we acknowledge technical support from S. Diewald, and L. Radtke. 
Funding was provided by the Deutsche Forschungsgemeinschaft (DFG – German Research Foundation) under project number 450396347 (GeHoldeQED).
Funding for the noise calibration was provided by the European Union under the Horizon Europe Program, grant agreement number 101080152 (TruePA).
I.T. and T.R. acknowledge support from the German Ministry of Education and Research (BMBF) within the project GEQCOS (FKZ: 13N15683).
A.N. acknowledges financing from the Baden-Württemberg Stiftung within the project QT-10 (QEDHiNet). S.Ge. and S.I. acknowledge support from  the European Commission (FET-Open AVaQus GA 899561). S.G., D.R. and W.W. acknowledge support from the Leibniz award WE 4458-5.
Facilities use was supported by the Karlsruhe Nano Micro Facility (KNMFi) and KIT Nanostructure Service Laboratory (NSL). We acknowledge qKit for providing a convenient measurement software framework.
\end{acknowledgments}

\appendix
\renewcommand{\appendixname}{Appendix}

\section*{Supplemental material}

\section{Inductive participation ratio of grAl nanostrip}
\label{A_Lkin}
The two main inductive contributions in each grAl resonator arise from the kinetic inductance of the grAl strip $L\mathrm{_{strip}}$, which constitutes the source of non-linearity in the grAlPA, and the inductance of the capacitor pads $L\mathrm{_{pads}}$, which gives a negligible contribution to non-linearity (cf Fig. 1(c) and (d) in the main text). The geometric inductance is two orders of magnitude smaller and we neglect it. To extract an estimation of the sheet inductance $L\mathrm{^k_\square}$ of grAl, we use Mattis-Bardeen formula \cite{Tinkham}

\begin{equation}
L\mathrm{^k_\square}=\frac{R\mathrm{_\square} \hbar}{\pi \Delta}\approx 0.12 \,\mathrm{nH/\square},
\label{eq_Mattis-Bardeen}
\end{equation}
where $R\mathrm{_\square}$ = $\rho/t$ is the sheet resistance and $\rho$ the resistivity of the grAl film, $t$ is the thickness of the film and $\Delta$ is its superconducting gap estimated from Ref.~\cite{Levy-Bertrand2019}. The grAl strip accounts for 35~squares of sheet inductance, which gives $L\mathrm{_{strip}}$ = 4.20 nH. Using finite element method (FEM) simulations we obtain $L\mathrm{_{pads}} \approx$~1~nH. Consequently, the inductance participation ratio of the grAl central line $p$~=~$L\mathrm{_{strip}}/(L\mathrm{_{strip}}+L\mathrm{_{pads}})$ is approximately~80\%. 

\begin{figure}[t]
\begin{center}
\includegraphics[width = 1\columnwidth]{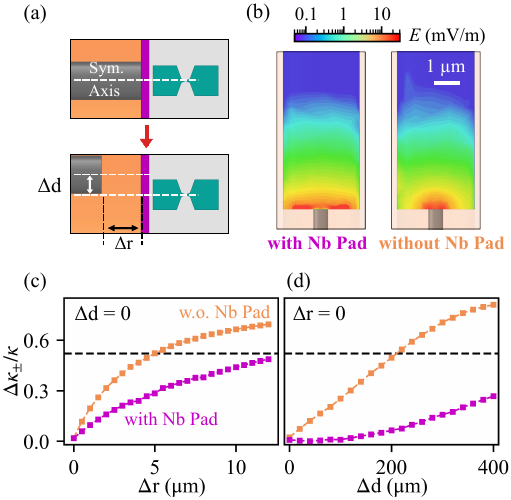}
\caption{\textbf{Finite element simulations of the hybridization susceptibility to misalignments.} \textbf{(a)}~Schematic representation of the two cases considered: thermal retraction $\Delta\mathrm{r}$ of the coaxial pin and lateral displacement $\Delta\mathrm{d}$ with respect to the symmetry axis of the waveguide (indicated by the white dashed line). The color code is the same as in Fig. 1(c) in the main text. \textbf{(b)}~Comparison of the electric field distribution across the surface of a sapphire chip with and without the Nb pad. \textbf{(c)},~\textbf{(d)}~Coupling asymmetry $\Delta\kappa\mathrm{_{\pm}}/\kappa$ as a function of $\Delta\mathrm{d}$ and $\Delta\mathrm{r}$, respectively. The black dashed line represents an empirical limit above which non-degenerate gain is not achievable (cf. Table \ref{tab:LinearCharact}).}
\label{Nb_Pad}
\end{center}
\end{figure}

\begin{table*}[t]
\caption{Summary of the linear characterization of grAlPAs with and without Nb coupling pad. The parameters were extracted from independent cooldowns using the same sample holder. The uncertainty in $\gamma\mathrm{_{\pm}}$ arises from the Fano uncertainty in the measurement setup \cite{Rieger2023fano}. We also estimate 10$\%$ uncertainty in $\kappa\mathrm{_{\pm}}$ due to the same effect. }
\begin{center}
\begin{tabular}{  p{3.2cm} | p{1.89cm} p{1.89cm} p{1.89cm} p{1.89cm} | p{1.89cm} p{1.89cm} p{1.89cm}  } 
\hline 
 & \multicolumn{4}{ c|}{\cellcolor{wNbPad!40} {with Nb Pad}} & \multicolumn{3}{c}{\cellcolor{woNbPad!45} {without Nb Pad}} \Tstruttop\Bstruttop  \\ \hline
\hfil Cooldown $\#$ & \hfil 1 & \hfil 2 & \hfil 3 & \hfil 4 & \hfil 1 & \hfil 2 & \hfil 3 \Tstrut\Bstrut \\ \hline
\hfil $\omega\mathrm{_+}/2\pi$ (GHz) & \hfil 8.41 &\hfil 8.38 & \hfil8.42 & \hfil8.40 &\hfil 9.35 & \hfil9.67 &\hfil 9.56 \Tstrut\Bstrut \\ 
\hfil $\omega\mathrm{_-}/2\pi$ (GHz) & \hfil8.21 & \hfil8.18 &\hfil 8.22 & \hfil8.22 & \hfil 8.38 &\hfil  8.92 & \hfil8.90 \Tstrut\Bstrut \\ 
\hfil $\kappa\mathrm{_+}/2\pi$ (MHz) & \hfil23.1 & \hfil 28.2 & \hfil28.4 & \hfil26.1 & \hfil91.9 & \hfil 5.8  & \hfil 85.0 \Tstrut\Bstrut \\
\hfil $\kappa\mathrm{_-}/2\pi$  (MHz) & \hfil 34.6 & \hfil 38.4 & \hfil 28.1 & \hfil 19.9 & \hfil 15.6 & \hfil 26.1 &\hfil  29.5 \Tstrut\Bstrut \\
\hfil $\gamma\mathrm{_+}/2\pi$ (MHz) & \hfil $\leq$ 5.0 & \hfil $\leq$ 8.0 & \hfil $\leq$ 7.4  &  \hfil $\leq$ 6.3 &\hfil  $\leq$ 20.6 & \hfil $\leq$ 1.3 & \hfil $\leq$ 21.1 \Tstrut\Bstrut \\ 
\hfil $\gamma\mathrm{_-}/2\pi$ (MHz) & \hfil $\leq$ 6.7 & \hfil $\leq$ 10.9 & \hfil $\leq$ 6.8  & \hfil $\leq$ 4.4 &\hfil $\leq$ 2.8 & \hfil $\leq$ 5.4 & \hfil $\leq$ 6.5 \Tstrut\Bstrut \\
\hfil $\omega\mathrm{_L}/2\pi$ (GHz) &\hfil 8.29 & \hfil8.27 &\hfil 8.32  &\hfil 8.33 &\hfil 9.21 & \hfil9.06 & \hfil9.39 \Tstrut\Bstrut \\
\hfil $\omega\mathrm{_R}/2\pi$ (GHz) &\hfil 8.33 &\hfil 8.30 &\hfil 8.32 & \hfil8.33 & \hfil 8.52 & \hfil 9.53 & \hfil 9.07 \Tstrut\Bstrut \\
\hfil $J/2\pi$ (MHz) &\hfil  99.0 &\hfil 97.0 & \hfil 99.0  & \hfil 90.0 & \hfil 341.0 & \hfil 286.0 &\hfil 286.0 \Tstrut\Bstrut \\
\hfil $\kappa/2\pi$ (MHz) &\hfil 57.7 & \hfil 66.6 &\hfil 56.5  & \hfil 46.0 & \hfil 107.5 & \hfil 31.9 &\hfil  114.5 \Tstrut\Bstrut \\ 
\hfil $\Delta \kappa\mathrm{_{\pm}}/\kappa$ & \hfil 0.20 & \hfil 0.15 &\hfil  0.01  & \hfil 0.14 & \hfil 0.71 & \hfil 0.64 & \hfil 0.48 \Tstrut\Bstrut \\
\hfil max. $G\mathrm{_0}$ (dB) & \hfil $\geq$ 20 & \hfil $\geq$ 20 & \hfil $\geq$ 20 & \hfil $\geq$ 20 &\hfil  None & \hfil None &\hfil  12 \Tstrut\Bstrut \\ \hline
\end{tabular}
\end{center}
\label{tab:LinearCharact}
\end{table*}

\section{Fabrication}
\label{A_FabDes}

The grAlPA is fabricated using a single step of electron beam lithography on top of a c-plane sapphire wafer with thickness 330~\SI{}{\micro \meter}. The substrate is coated with a resist stack of 800~nm MMA EL-13 and 400~nm PMMA A4, followed by an additional 10~nm Au antistatic layer. The grAl resonators are then patterned with a 50~keV e-beam writter and developed using an IPA:$\mathrm{H_2}$O (3:1) solution at 6$^{\circ}$C. Before the metal deposition, the substrate is cleaned with a Ar/$\mathrm{O_2}$ descum process inside a Plassys electron beam evaporator. To improve the vacuum conditions after the initial cleaning step, we use Ti as a getter material. A grAl layer with thickness $t$~=~40~nm and resistivity $\rho$~=~830~\SI{}{\micro \ohm}$\,$cm is deposited under zero angle Al evaporation, at 1~nm/s deposition rate and using dynamical oxidation. Finally, the sample is oxidized in-situ for 3~min in a fixed $\mathrm{O_2}$ pressure of 10~mbar, to prevent degradation of the grAl film after exposure to ambient conditions. The Nb coupling pad visible in Fig. 1(c) in the main text is fabricated in a subsequent optical lithography step using a mask aligner. For the Nb deposition, we use the same substrate cleaning and Ti gettering as for the grAl resonators. We evaporate a 40~nm Nb film at a deposition rate of about $1.2$~nm/s. 

\section{Engineering the hybridization of the grAl resonators}
\label{A_LinCharac}

\subsection{Hybridization and coupling}
\label{A_Dimers}
We now turn our attention to the impact of the grAl resonator hybridization on the gain performance of the grAlPA. By linearizing the Hamiltonian in Eq. (1) of the main text, assuming negligible internal losses ($\kappa\mathrm{_{\pm}}$) and following the approach of Refs. \cite{Eichler2014Dimer,IvanThesis2022}, we derive

\begin{equation}
\omega\mathrm{_\pm}=\frac{\omega\mathrm{_L}+\omega\mathrm{_R}}{2} \pm \sqrt{J^2 + \left(\frac{\omega\mathrm{_L}-\omega\mathrm{_R}}{2}\right)^2} 
\label{eq_omegas_hybrid}
\end{equation}

and

\begin{equation}
\kappa\mathrm{_\pm}=\frac{\kappa}{2}\left( 1 \pm \frac{\omega\mathrm{_L}-\omega\mathrm{_R}}{\sqrt{4J^2 + \left(\omega\mathrm{_L}-\omega\mathrm{_R}\right)^2}} \right),
\label{eq_kappas_omegas_hybrid}
\end{equation}
where $\kappa\mathrm{_\pm}$ are the linewidths of the dimer modes. The scaling of the grAlPA instantaneous bandwidth (BW) is given by $\sqrt{G\mathrm{_0}}\cdot\mathrm{BW}$~=~$\kappa\mathrm{_{eq}}$, where $G\mathrm{_0}$ is the maximum gain of the grAlPA and the equivalent linewidth $\kappa\mathrm{_{eq}}$ is defined as
\begin{equation}
\kappa\mathrm{_{eq}}=\frac{2 \kappa_+ \kappa_-}{\kappa_+ + \kappa_-}.
\label{eq_kappas_equivalent}
\end{equation}

Perfect hybridization, $\omega\mathrm{_L}~=~\omega\mathrm{_R}$, gives the maximum BW and the total coupling strength $\kappa$ is equally split between the dimer modes, $\kappa\mathrm{_+}$~=~$\kappa\mathrm{_-}$~=~$\kappa$/2. If $\omega\mathrm{_L} \neq \omega\mathrm{_R}$, besides the reduction of the BW of the amplifier, following \cref{eq_kappas_omegas_hybrid}, the linewidths of the dimer modes also show an asymmetry $\Delta \kappa_\mathrm{_\pm}$~=~$|\kappa\mathrm{_+}-\kappa\mathrm{_-}|$. Consequently, we use the asymmetry $\Delta \kappa_\mathrm{_\pm}$ as a proxy for the hybridization of the resonators.

\subsection{The role of the Nb pad}
\label{A_Nb_Pad}

Figure \ref{Nb_Pad} illustrates that the Nb pad reduces the hybridization susceptibility to misalignments. We consider two possible misalignments: lateral displacements ($\Delta d$) due to imperfections of the substrate, and thermal contraction ($\Delta r$) of the coaxial pin during the cooldown (see \cref{Nb_Pad}(a)). In \cref{Nb_Pad}(b) we plot the on-chip distribution of the electric field obtained from finite element simulations in the absence of grAl resonators. We use a driving excitation at 8~GHz and an input power equivalent to a dimer population of 10 photons. We observe that the addition of the Nb pad spreads the electric field lines along the lower edge of the substrate. As a consequence, when using the Nb pads the coupling asymmetry $\Delta\kappa\mathrm{_\pm}$ obtained from simulations is less susceptible to misalignments, as shown in \cref{Nb_Pad}(c) and (d). 

In Table \ref{tab:LinearCharact} we compare two grAlPAs fabricated with and without a Nb pad, measured in several successive cooldowns. As expected from simulations, the device with a Nb pad consistently provides gain $\geq$ 20~dB in all four cooldowns, while the sample without a coupling pad only shows non-degenerate gain in one of three cooldowns. For both devices the circuit parameters are extracted from single-tone spectroscopy using \cref{eq_omegas_hybrid,eq_kappas_omegas_hybrid}.

\section{Kerr coefficients}
\label{A_Non_linearities}

The non-linearity of our grAl resonators stems from the central grAl strip shown in Fig. 1(d) of the main text. Following Refs. \cite{Maleeva2018, IvanThesis2022}, we estimate the value of the Kerr coefficient $K$ by modelling the strip as an effective array of $N\mathrm{_{J}}$ Josephson Junctions (JJ). We consider that the critical current of each junction $I\mathrm{_c^{J}}$ equals the critical current of the grAl strip $I\mathrm{_c^{grAl}}$ and provides a Josephson inductance $L\mathrm{_{J}}$~=~$\Phi\mathrm{_0}/2\pi I\mathrm{_c^{J}}$, where $\Phi\mathrm{_0}$~=~$h/2e$ denotes the magnetic flux quantum. The inductance of the grAl strip is $L\mathrm{_{strip}}$~=~$N\mathrm{_{J}} L\mathrm{_{J}}$ and its corresponding self-Kerr coefficient is

\begin{equation}
K=-\frac{E\mathrm{_{J}}}{8\hbar N\mathrm{_{J}^3}} \left( \frac{\omega\mathrm{_{r}} L\mathrm{_{strip}}}{R\mathrm{_Q}}\right)^2,
\label{eq_theory_Kerr}
\end{equation}
where $E\mathrm{_{J}}$ = $\Phi\mathrm{_0} I\mathrm{_c^{J}}/2\pi $ is the Josephson energy of the equivalent JJs, $\omega\mathrm{_{r}}$ the frequency of the resonator and $R\mathrm{_Q}$~=~$\hbar/4e^2$ the resistance quantum. We estimate $I\mathrm{_c^{grAl}}$ from the resistivity of the film, as shown in Ref. \cite{IvanThesis2022}, while the rest of the parameters are extracted from Table \ref{tab:LinearCharact}. For the grAlPA, the JJ array model yields $K/2\pi$~=~6-7~kHz.

Experimentally, we measure the non-linearity from the power-dependent frequency shifts of the dimer modes. Assuming equal Kerr coefficients for both grAl resonators $K\mathrm{_L} \approx K\mathrm{_R}$ = $K$ and applying the rotating-wave-approximation, the non-linear part of the grAlPA Hamiltonian can be rewritten as

\begin{equation}
H\mathrm{_{NL}}/\hbar = \sum_{i= +,-}\frac{K}{2}a^{\dagger}_{i}a_{i}a^{\dagger}_{i}a_{i} + \sum_{i < j}\frac{K}{2}a^{\dagger}_{j}a_{j}a^{\dagger}_{i}a_{i},
\label{eq_Dressed hamiltonian}
\end{equation}
where the terms on the right hand side of the equation give the self-Kerr and cross-Kerr frequency shifts, respectively. The mode population $\overline{n}\mathrm{_\pm}$ increases linearly with input power $P\mathrm{_{in}}$, following

\begin{equation}
\overline{n}\mathrm{_\pm} = \frac{4 P\mathrm{_{in}}}{\hbar \omega\mathrm{_{\pm}}} \frac{\kappa\mathrm{_{\pm}}}{\left( \kappa\mathrm{_{\pm}} + \gamma\mathrm{_{\pm}} \right)^2},
\label{eq_population}
\end{equation}
where $\omega\mathrm{_{\pm}}$ is the resonant frequency of each dimer mode and $\kappa\mathrm{_{\pm}}$, $\gamma\mathrm{_{\pm}}$ their external and internal coupling parameters, respectively. The power at the input of the grAlPA ($P\mathrm{_{in}}$) is estimated using the calibration shown in Appendix \ref{A_Magnetometer}. As presented in \cref{Kerrs_plot}, all Kerr coefficients of the grAlPA are measured to be in the range of 1-3 kHz and they are unaffected by in-plane magnetic field up to 1~T.

\begin{figure}[!t]
\begin{center}
\includegraphics[width = 1\columnwidth]{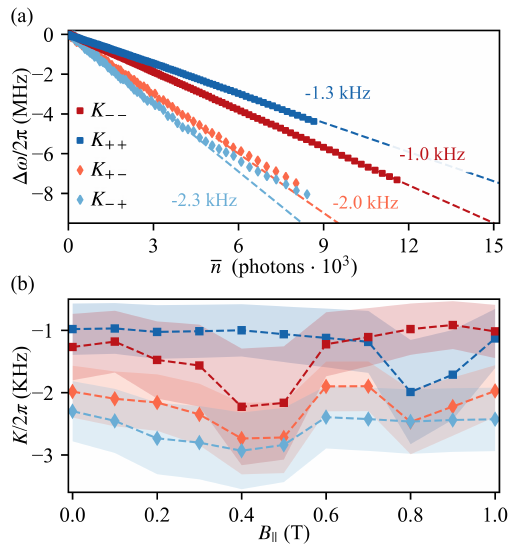}
\caption{\textbf{Self-Kerr and cross-Kerr non-linearities.} \textbf{(a)}~Shift in the resonance frequency of each grAlPA mode as a function of their average photon number. The shift produced by the self-Kerr ($K_{++}$ and $K_{--}$) and cross-Kerr ($K_{+-}$ and $K_{-+}$) non-linearities are measured through single-tone and two-tone spectroscopy, respectively. \textbf{(b)}~Magnetic field dependence of the Kerr coefficients. Note that all coefficients remain within the range of 1-3~kHz up to $B_\mathrm{||}$~=~1~T.}
\label{Kerrs_plot}
\end{center}
\end{figure}

\section{Internal losses}
\label{A_Losses}

\begin{figure}[!t]
\begin{center}
\includegraphics[width = 1\columnwidth]{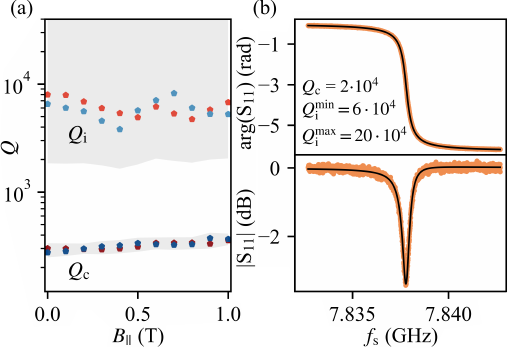}
\caption{\textbf{Internal and external quality factors.} \textbf{(a)}~Field dependence of the internal and external quality factors for single-photon population. The size in the error region of $Q\mathrm{_i}$ (upper grey area) comes from the Fano uncertainty of the measurement and the fact that each mode of the grAlPA is overcoupled~\cite{Rieger2023fano}. Note that $Q\mathrm{_i}$ is at least 2 orders of magnitude above $Q\mathrm{_c}$ for the entire field range. \textbf{(b)}~Single-tone spectroscopy of a sample with critical coupling $Q\mathrm{_i} \approx Q\mathrm{_c}$. For this case a reduced Fano uncertainty allows a more accurate estimation of the expected $Q\mathrm{_i}$. The data in panel (b) is adapted from \cite{IvanThesis2022} with permission.}
\label{Qis_plot}
\end{center}
\end{figure}

Here we report the magnetic field dependence of the internal losses in the grAlPA. We calculate the single-photon internal quality factor $Q\mathrm{_{i}}$ and external quality factor $Q\mathrm{_{c}}$ of each dimer mode from circle fits \cite{Probst2015Feb} of their respective reflection coefficients at different $B_\mathrm{||}$. As illustrated in \cref{Qis_plot}(a), both modes preserve $Q\mathrm{_{i}}/Q\mathrm{_{c}} \gtrsim$~10 up to $B_\mathrm{||}$ = 1~T. Due to the strong coupling required for the operation of the amplifier, we can only place a lower bound for the $Q\mathrm{_{i}}$ value~\cite{Rieger2023fano}. To extract a more accurate estimation of the internal losses, we would need to decouple the resonators such that they are critically coupled i.e. $Q\mathrm{_{i}} \approx Q\mathrm{_{c}} $. In \cref{Qis_plot}(b) we show the resonant response of one dimer mode for a critically coupled grAlPA, from which we obtain $Q\mathrm{_{i}}$~=~9$\cdot 10^4$ with the Fano uncertainty range 6$\cdot 10^4$~$<$~$Q\mathrm{_{i}}$~$<$~20$\cdot10^4$.

\section{Magnetic field dependence of the frequency and coupling of the grAl resonators}
\label{A_Decoupling}
The frequency of the grAl resonators experiences a shift when we apply an in-plane magnetic field $B_\mathrm{||}$. This effect is produced by the suppression of the superconducting gap $\Delta$ in field \cite{Winkel2020Transmon},

\begin{figure}[!t]
\begin{center}
\includegraphics[width = 1\columnwidth]{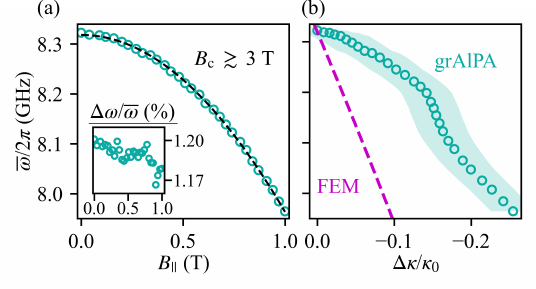}
\caption{\textbf{Magnetic field dependence of $\overline{\omega}$ and $\kappa$.} \textbf{(a)}~Field dependence of the average frequency $\overline{\omega}$~=~($\omega\mathrm{_{L}}+\omega\mathrm{_{R}}$)/2. The green circles are calculated from single-tone spectroscopy measurements and using \cref{eq_omegas_hybrid,eq_kappas_omegas_hybrid}. The dashed black line shows the fit using \cref{eq_Delta_omega_field}, from which we extract a critical field $B\mathrm{_c} \gtrsim$~3~T. The inset displays the relative frequency difference between the modes $\Delta\omega$/$\overline{\omega}$, with $\Delta\omega$~=~$\omega\mathrm{_{R}}-\omega\mathrm{_{L}}$. \textbf{(b)}~Change of the total coupling $\Delta \kappa$ as a function of $\overline{\omega}$. The shaded region represents the Fano uncertainty in the estimation of the external coupling strength $\kappa$ \cite{Rieger2023fano}. The purple dashed line is obtained from finite element simulations. The values of $\Delta \kappa$ are normalized relative to the coupling $\kappa\mathrm{_0}$ at $B_\mathrm{||}$~=~0~T.}
\label{Coupling_Bare_Field}
\end{center}
\end{figure}

\begin{equation}
\Delta(B_\mathrm{||})=\Delta(\mathrm{0})\sqrt{\frac{1-(B_\mathrm{||}/B \mathrm{_c})^2}{1+(B_\mathrm{||}/B \mathrm{_c})^2}},
\label{eq_Delta_Field}
\end{equation}
where $B \mathrm{_c}$ is the in-plane critical field of grAl. From \cref{eq_Mattis-Bardeen} we expect an increase of the total kinetic inductance $L\mathrm{_k}$ and a frequency shift following

\begin{equation}
\omega(B_\mathrm{||})=\frac{1}{\sqrt{L\mathrm{_k}(B_\mathrm{||})C}}=\omega(0)\left(\frac{1-(B_\mathrm{||}/B \mathrm{_c})^2}{1+(B_\mathrm{||}/B \mathrm{_c})^2}\right)^{1/4},
\label{eq_Delta_omega_field}
\end{equation}
where $C$ is the resonator capacitance. To extract $B\mathrm{_c}$, we calculate the frequency shift of the grAl resonators from the measured magnetic field dependence of $\omega\mathrm{_\pm}$ shown in Fig. 1(f) of the main text and using \cref{eq_omegas_hybrid,eq_kappas_omegas_hybrid}. As presented in \cref{Coupling_Bare_Field}(a), we obtain $B \mathrm{_c} \approx$~3~T. A possible explanation of the fact that $B \mathrm{_c}$ in our experiments is lower than previously reported values of up-to 6~T \cite{Borisov2020} is the imperfect compensation of out-of-plane magnetic field $B_\mathrm{\perp}$ (for details see Appendix \ref{A_Field_Alignment}). Lastly, note that the frequencies of the grAl resonators differ by less than 1.5\% for all $B_\mathrm{||}$ up to 1~T, as shown in the inset of \cref{Coupling_Bare_Field}(a).

\begin{figure}[!b]
\begin{center}
\includegraphics[width = 1\columnwidth]{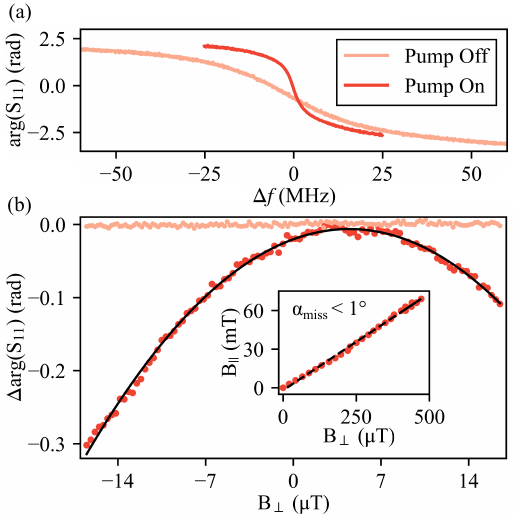}
\caption{\textbf{Compensation of residual out-of-plane magnetic field components.} \textbf{(a)}~Phase of the reflection coefficient $\mathrm{S_{11}}$ as a function of the detuning $\mathrm{\Delta} f$ from the resonance frequency $f\mathrm{_r}$, while having the grAlPA on and off. The narrowing of the response appears as a consequence of the gain-bandwidth trade-off in the grAlPA. We use a pump tone generating a maximum non-degenerate gain of $G\mathrm{_0}$~=~18~dB. This results in a factor of 8 decrease in the linewidth, which improves our detection sensitivity. \textbf{(b)}~Example of a compensation sweep employed to calibrate the unwanted $B\mathrm{_\perp}$, while fixing $B\mathrm{_{||}}$ at 0~T. We sweep $B\mathrm{_\perp}$ and monitor the change in the phase response $\Delta$arg(S$\mathrm{_{11}}$) for a fixed frequency close to $f\mathrm{_r}$. Note that a non-negligible variation is observed only when the grAlPA pump is on. We determine the compensation field $B\mathrm{_\perp}$ by fitting the data with a quadratic function (black solid line) and extracting the field that maximizes $\Delta$arg(S$\mathrm{_{11}}$). The inset displays the compensation fields measured for several values of $B\mathrm{_{||}}$. The black dashed line depicts a linear fit of the data. The slope of this line provides an estimation of the misalignment angle of the in-plane component, which corresponds to an angle less than 1$\mathrm{^{\circ}}$.}
\label{field_Alignment}
\end{center}
\end{figure}

The decrease of external coupling $\kappa$ with in-plane field is depicted in \cref{Coupling_Bare_Field}(b) and is also visible in the reduction of the instantaneous bandwidth of the grAlPA (see Fig. 3(a) in the main text). It arises from the gradual decoupling of the dimer modes as their frequencies move farther from the 60~GHz cutoff of the cylindrical waveguide, as qualitatively confirmed by FEM simulations. We attribute the difference between simulations and measurements of $\kappa$ to additional features in the environmental impedance seen by the grAlPA, which could be related to the frequency-dependent isolation of the circulators used in our setup~\cite{Rieger2023fano}. 

\section{Magnetic field alignment to minimize the out-of-plane component}
\label{A_Field_Alignment}

The residual out-of-plane magnetic field component $B\mathrm{_{\perp}}$ from misalignment of the substrate inside the cylindrical sample holder in Fig. 1 of the main text can lead to trapping of vortices in the grAl resonators \cite{Bothner2011, Kroll2019}. Trapped vortices produce additional loss channels potentially adding noise to the amplifier, especially for $B\mathrm{_{||}}$ approaching 1~T. To mitigate this issue, we calibrate the orientation of the magnetic field following a similar procedure as explained in Ref. \cite{Borisov2020}. We fix $B\mathrm{_{||}}$ and sweep $B\mathrm{_\perp}$ in a range small enough to avoid vortex trapping. Simultaneously, we monitor the phase response in one of the dimer modes at a fixed frequency close to resonance. The phase then follows a quadratic variation with $B\mathrm{_\perp}$, where the maximum indicates the required compensation field. 

Even though the 200~nm wide central grAl strip (cf. Fig. 1(d) and (e) in the main text) provides out-of-plane magnetic field resilience in the range of 0.1~T~\cite{Janik2024Jul}, the field range we can use for the compensation sweep before we trap flux is limited by the width of the capacitor pads~\cite{Stan2004Mar}, which restricts us to a field range almost two orders of magnitude lower compared to Refs.~\cite{Borisov2020,Janik2024Jul}. This limitation is exacerbated by the fact that the modes of the amplifier are strongly coupled, therefore their phase susceptibility to $B\mathrm{_{\perp}}$ is relatively low. To relax this limitation we take advantage of the grAlPA gain-bandwidth product. By applying a pump tone, the linewidths of the dimer modes decrease according to $\sqrt{G\mathrm{_0}}\cdot\mathrm{BW}$~=~$\kappa\mathrm{_{eq}}$ (see \cref{eq_kappas_equivalent}), which for $G\mathrm{_0}$~=~18~dB effectively reduces the linewidth of the dimer modes by a factor of 8 (see \cref{field_Alignment}(a)). We therefore enhance the sensitivity of the phase to $B\mathrm{_\perp}$, facilitating the measurement of the compensation field. \cref{field_Alignment}(b) illustrates the result of a compensation sweep at $B\mathrm{_{||}}$~=~0~T for the low-frequency dimer mode, with the pump on and off. We observe a clear quadratic trend in the phase only when the pump is on. By repeating the same procedure for different in-plane fields, we obtain the dependence of the compensation field with $B\mathrm{_{||}}$, which follows a linear trend as depicted in the inset of \cref{field_Alignment}(b). From the slope of this linear response we extract a misalignment angle of the in-plane magnetic field $\alpha\mathrm{_{miss}} <$ 1$\mathrm{^{\circ}}$.

\begin{figure}[!t]
\begin{center}
\includegraphics[width = 1\columnwidth]{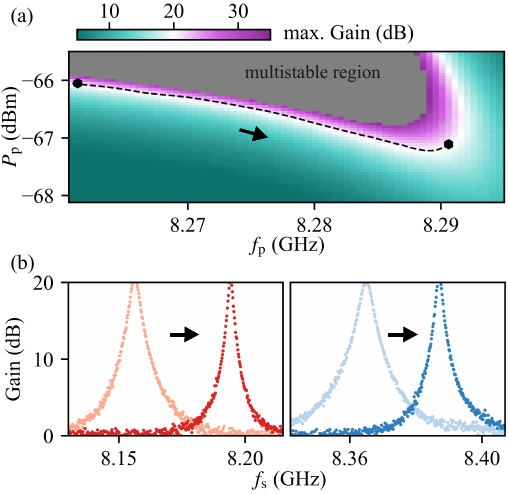}
\caption{\textbf{Frequency tunability of the gain profile.} \textbf{(a)}~Maximum gain of grAlPA as a function of the pump frequency and pump power. Regions with 20 dB maximum gain are highlighted by a white fringe, while the grey area represents the multistable regime of the amplifier. The gain profile of the grAlPA is tuned by moving the pump tone along the dashed black line, as shown in \textbf{(b)}. We achieve frequency tunability close to 50~MHz for each dimer mode. The narrowing of the Lorentzian gain profiles is attributed to variations in the environmental impedance seen by the grAlPA.}
\label{3D_Gain}
\end{center}
\end{figure}

\begin{figure*}[t!]
\begin{center}
\includegraphics[width=6.67in]{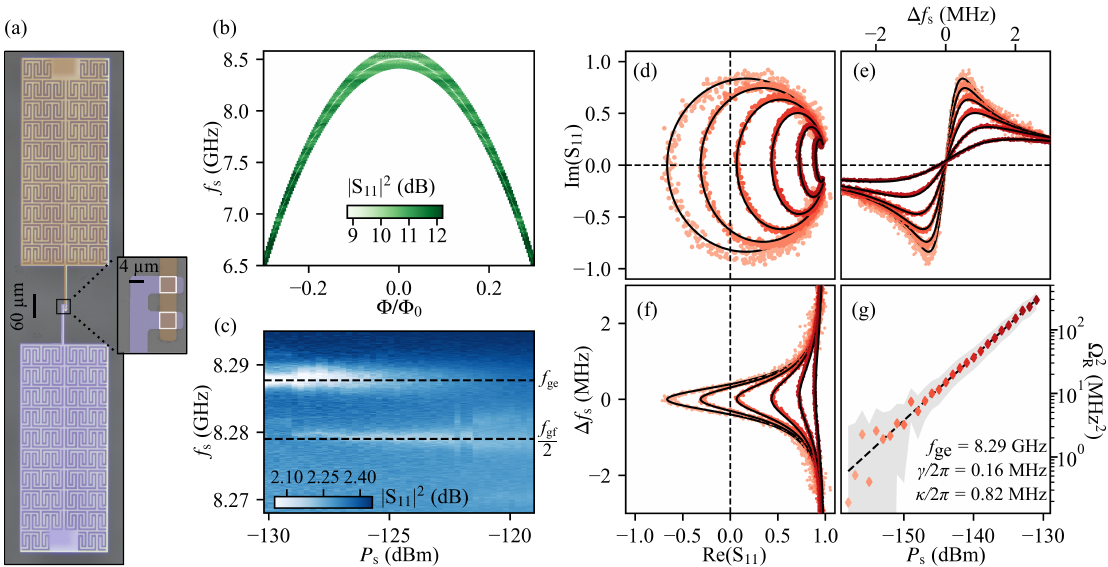}
\caption{\textbf{Resonance fluorescence of a transmon qubit.} \textbf{(a)}~False-colored image of the flux tunable transmon qubit used for power calibration. The device is fabricated using 2 steps of optical lithography (brown and purple layers) similar to the implementations shown in \cite{GunzlerMagnetometer, Winkel2020DJA}. In the inset, we show a magnified image of the squid loop with each Josephson junction highlighted by a white square. \textbf{(b)}~Flux dependence of the spectrum of the transmon qubit. The resonance frequency of the qubit appears as a dip of the reflection amplitude $|\mathrm{S_{11}}|$. \textbf{(c)}~Single-tone spectroscopy of the transmon qubit as a function of on-chip power $P\mathrm{_{s}}$. Form the first ($f\mathrm{_{ge}}$) and second transition ($f\mathrm{_{gf}}/2$) frequencies, we extract an anharmonicity $\alpha \approx$~18~MHz. \textbf{(d)}~Quadrature plane of the reflection coefficient in the vicinity of the qubit frequency at $f\mathrm{_{ge}}$=~8.287~GHz. At low powers the frequency follows a circular shape similar to the response of a linear resonator, from which we extract $\kappa$/$2\pi$~=~828~KHz and $\gamma$/$2\pi$~=~155~KHz through a standard circle fit procedure \cite{Probst2015Feb}. As power increases, the shape transforms into an ellipse, characteristic of the resonance fluorescence of a two-level system \cite{GunzlerMagnetometer, Winkel2020Transmon,IvanThesis2022,Astafiev2010Fluoresnce}.
\textbf{(e)},~\textbf{(f)}~Real and imaginary part of the reflection coefficient as a function of the detuning from the qubit frequency $\Delta f\mathrm{_s}$. The black solid lines in panels d), e) and f) are fits of the measurement data using equation (\ref{eq_FluorescenceS11}). For the fits, we fix the values of $\kappa$, $\gamma$ and $f\mathrm{_{ge}}$ calculated at the lowest power and use $\Omega\mathrm{_R}$ as the sole fitting parameter. \textbf{(g)}~Input power dependence of the Rabi frequency $\Omega\mathrm{_R}$. The shaded area indicates the Fano uncertainty in $\kappa$ and $\gamma$.  We fit the data using equation (\ref{eq_RabiAttenuation}), from which we extract an input attenuation of -89.3~$\mathrm{\pm}$~2~dB.}
\label{Magnetometer_plot}
\end{center}
\end{figure*}

\section{Tunability of the gain profile with pump power and frequency}
\label{A_Frequency_Modulation}

The grAlPA offers additional gain tunability by varying the frequency $f\mathrm{_p}$ and power $P\mathrm{_p}$ of the pump tone. \cref{3D_Gain}(a) illustrates the dependence of the maximum gain $G\mathrm{_0}$ with $P\mathrm{_p}$ and $f\mathrm{_p}$. We calculate $G\mathrm{_0}$ by fitting the gain profiles with a Lorentzian function. As marked by the white regions in \cref{3D_Gain}(a), it is possible to obtain 20~dB gain for various combinations of $P\mathrm{_p}$ and $f\mathrm{_p}$. Moreover, by increasing the pump power it is possible to increase the gain up to 30~dB, as also depicted in \cref{3D_Gain}(a). For higher powers the grAlPA enters a multistable regime \cite{Eichler2014Dimer,IvanThesis2022}, where parametric amplification is not achievable using our pumping scheme. Note that moving across the 20~dB region, we can tune the gain profile of each dimer mode by approximately $\approx$ 50 MHz, as shown in \cref{3D_Gain}(b). The change in bandwidth of the gain profiles is attributed to the frequency dependence of the environmental impedance seen by the grAlPA, similar to the effect observed in impedance engineered parametric amplifiers \cite{Mutus2014Jun}.

\section{Power calibration using the resonant fluorescence of a transmon qubit}
\label{A_Magnetometer}

\subsection{Transmon qubit}
\label{A_TransmonQubit}

To characterize the noise performance in our amplifiers we use the resonant fluorescence produced by the transmon qubit depicted in \cref{Magnetometer_plot}(a). We use a similar geometry as in Ref \cite{GunzlerMagnetometer}. The qubit is designed to fit the cylindrical sample holder used for the grAlPA and consists of a single SQUID junction shunted by a coplanar capacitor with $C\mathrm{_{p}}$~=~77~fF. The capacitor plates are fabricated using a fractal geometry to prevent unwanted flux-trapping while applying an out-of-plane magnetic field. We fabricate the qubit using two steps of optical lithography, followed by zero-angle evaporations of 30~nm and 40~nm pure Al, respectively. Before the evaporation of the second layer, we remove the native oxide layer of the first film employing an Ar milling procedure \cite{Grunhaupt2017Aug}, followed by a static oxidation for 30~min with an oxygen partial pressure of 30~mbar. 

The Josephson junctions (JJs) are formed from the overlap of the two Al layers, having an approximate area of 3.8x3.8~\SI{}{\micro \meter}$\mathrm{^2}$. As highlighted in the inset of \cref{Magnetometer_plot}(a), the SQUID loop in our design has an area about 4x2~\SI{}{\micro \metre}$\mathrm{^2}$. From the geometry of the JJs we expect a junction capacitance $C\mathrm{_{J}}$~=722~fF and total SQUID capacitance $C\mathrm{_{SQ}}$~=~2$C\mathrm{_{J}}$~=1444~fF. Considering the contribution from the capacitor plates, we expect a total charging energy $E\mathrm{_C}/2\mathrm{\pi}$~=~13~MHz. We calculate the Josephson energy $E\mathrm{_J}/2\mathrm{\pi}$ from the qubit frequency $f\mathrm{_{ge}}$~=~$\sqrt{8E\mathrm{_C}E\mathrm{_J}}-E\mathrm{_C}$, measured using single-tone spectroscopy. We obtain $E\mathrm{_J}/2\mathrm{\pi}$~=~712~GHz, which confirms that our qubit is in the transmon regime. By increasing the drive power of the qubit we observe the appearance of the second transition frequency $f\mathrm{_{gf}}/2$, from which we calculate an anharmonicity of $\alpha \approx$~18~MHz. The field dependence of the qubit frequency is presented in \cref{Magnetometer_plot}(b), which shows that we can operate the qubit within the same frequency range as the grAlPA and use it for power calibration.

\subsection{Power calibration}
\label{A_powerCal}

\begin{figure}[!t]
\begin{center}
\includegraphics[width = 1\columnwidth]{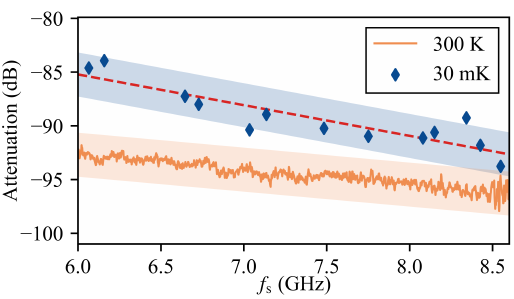}
\caption{\textbf{Input power calibration of the measurement setup.} The scattered data in blue represent the low-temperature attenuation of the input lines (see complete setup in Appendix \ref{A_SetUp}). The attenuation is calibrated using the resonance fluorescence of a transmon qubit. The red dashed line denotes the linear fit applied to estimate the low-temperature attenuation used in the characterization of the grAlPA. The error region comes from the uncertainty in the power calibration. For comparison, the orange line shows the attenuation of the input line measured at room temperature. The shaded area indicates the error range. }
\label{Attn_SionLudi}
\end{center}
\end{figure}

\begin{figure}[!t]
\begin{center}
\includegraphics[width = 1\columnwidth]{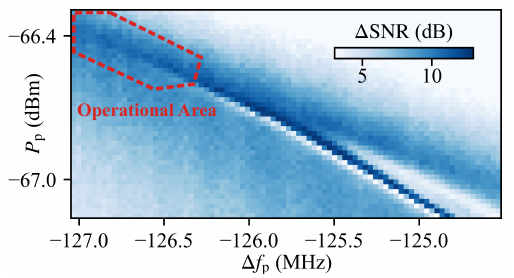}
\caption{\textbf{$\Delta$SNR as a function of pump power and signal-pump detuning $\Delta f\mathrm{_p}$ for a fixed coherent tone.} The operational area for the grAlPA is defined by the range of parameters where the device is stable and offers $\Delta$SNR larger than 10~dB, as indicated by the dashed contour in red.}
\label{SNR_3D}
\end{center}
\end{figure}

In \cref{Magnetometer_plot}(d)-(f) we show a typical transmon qubit fluorescence measurement. In the limit of low powers, close to the single-photon regime of the qubit, the shape of reflection coefficient in the quadrature plane approaches a circle, similar to a linear resonator. By increasing the driving power, we observe the characteristic elliptical shape coming from the fluorescence of a two-level system \cite{Winkel2020Transmon,GunzlerMagnetometer,Honigl-Decrinis2020Feb}. In the weak drive limit, when the on-resonance Rabi frequency of the qubit satisfies $\Omega\mathrm{_R}$~$\leq$~$\alpha$, the dependence of $S\mathrm{_{11}}$ with qubit-drive detuning $\Delta\mathrm{q}$ is given by

\begin{equation}
S\mathrm{_{11}} = 1-\frac{2\kappa}{\Gamma} \frac{1+2i\Delta\mathrm{_q}/\Gamma}{1+\left(2\Delta\mathrm{_q}/\Gamma\right)^2+2\left( \Omega\mathrm{_R}/\Gamma \right)^2},
\label{eq_FluorescenceS11}
\end{equation}
where $\Omega\mathrm{_R}$ is the on-resonance Rabi frequency of the qubit, $\kappa$ the external coupling rate, $\gamma$ the internal loss rate and $\Gamma$~=~$\kappa + \gamma$ the energy relaxation rate. We assume negligible pure dephasing rates $\Gamma\mathrm{_\phi}$, in comparison to $\Gamma$. For each driving power we fit the fluorescence of the transmon qubit using \cref{eq_FluorescenceS11}. The values of $\kappa/2\pi$~=~0.82~MHz and $\gamma/2\pi$~=~0.16~MHz are extracted from the fitting at powers close to the single-photon regime. For higher driving powers, we fix $\kappa$ and $\gamma$, and use $\Omega\mathrm{_R}$ as the only fitting parameter. 

We extract the attenuation $A$ in the input line of our setup by fitting the dependence of $\Omega\mathrm{_R}$ with driving power at room temperature $P\mathrm{_{RT}}$

\begin{equation}
\Omega\mathrm{^2_R} = A \frac{2\Gamma P\mathrm{_{RT}}}{h f\mathrm{_{ge}}},
\label{eq_RabiAttenuation}
\end{equation}
where the power reaching the input of the transmon qubit is given by $P\mathrm{_s} = A\cdot P\mathrm{_{RT}}$. As presented in \cref{Magnetometer_plot}(g), we observe a linear dependence between $P\mathrm{_s}$ and $\Omega\mathrm{^2_R}$, as expected from equation (\ref{eq_RabiAttenuation}). At 8.287~GHz we calculate an input attenuation of -89.3~$\pm$~2~dB. 

\begin{figure}[!t]
\begin{center}
\includegraphics[width = 1\columnwidth]{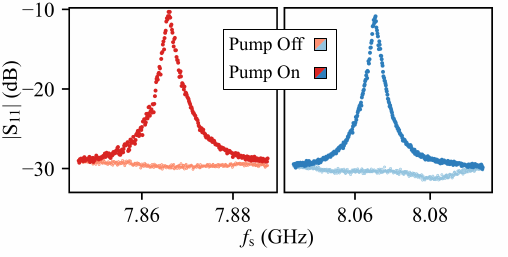}
\caption{\textbf{Reflection amplitude at $B \mathrm{_{||}}$~=~1~T.} We compare the amplitude of the reflection coefficient $\mathrm{S_{11}}$ close to each mode of the grAlPA, both with the pump off (light blue and red lines) and on (dark red and blue lines). When driving the grAlPA, the response resembles the expected Lorentzian gain profile. The dip observed at the resonance of the high frequency mode while having the pump off (light blue line on the right panel), explains the additional shoulder seen in the gain profile presented in Fig. 3(b) of the main text.}
\label{Gain_pumo_off}
\end{center}
\end{figure}

\begin{figure*}[!t]
\begin{center}
\includegraphics[width=6.67in]{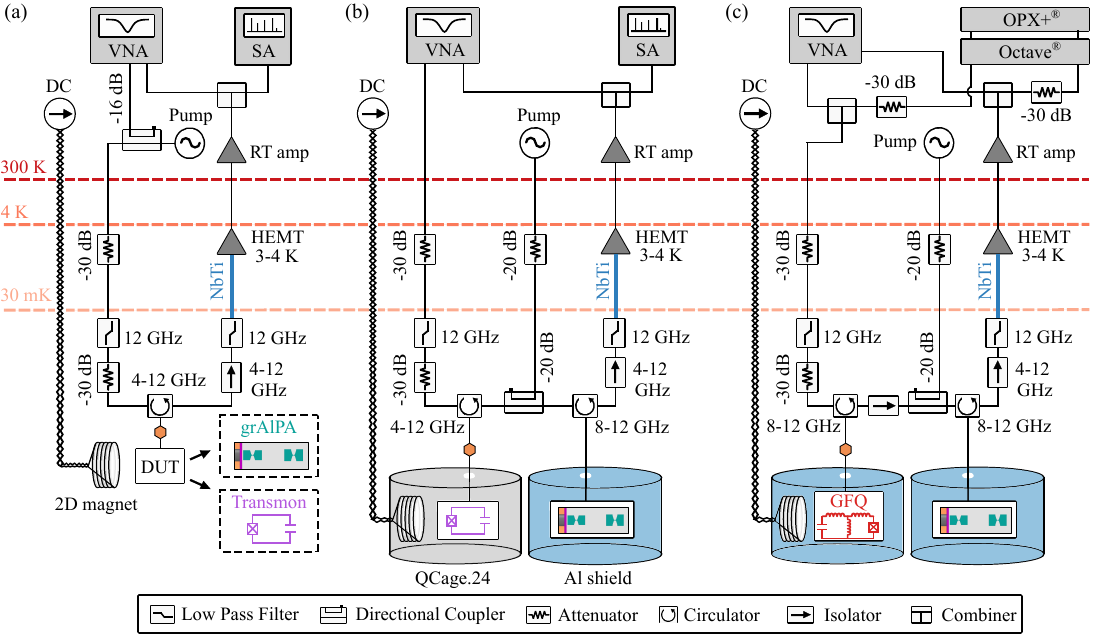}
\caption{\textbf{Experimental setups.} Each component is thermalized to the temperature stage indicated by the nearest dashed line. \textbf{(a)}~Schematics of the set up employed for the grAlPA measurements presented in the main text. We use the same magnetic-field biasing setup as in Refs. \cite{Borisov2020, Rieger2023, Rieger2023fano}. The magnetic field is generated by a homemade 2D vector magnet, with its orientation calibrated using the method described in Ref. \cite{Borisov2020} and in Appendix \ref{A_Field_Alignment}. To drive the grAlPA, the pump tone is combined through a directional coupler at room temperature with the signal from a vector network analyzer (VNA). A power divider located outside the dilution refrigerator splits the output signal between the receiving port of the VNA and a spectrum analyzer (SA). \textbf{(b)}~Setup employed for the measurements in Appendix~\ref{A_Noise_BF}. The grAlPA is connected via a circulator to a transmon qubit. The transmon qubit is housed inside a QCage.24 magnetic shielding \cite{Qcage} and the grAlPA inside a homemade Al shield. The pump tone is added using a cryogenic directional coupler. In this setup we estimate a total insertion loss between the transmon and the grAlPA on the order of 1.5 dB. \textbf{(c)} Setup employed for the GFQ measurements shown in Fig. 4(c)-(e) of the main text. The qubit is mounted inside a homemade Al shield. Compared to the setup in panel (b), we added an isolator and $\approx$~30~cm of microwave connections between the GFQ and the grAlPA, which add 0.8~dB insertion loss. The orange marker in all panels indicates the input power calibration plane.}
\label{SetUp}
\end{center}
\end{figure*}

The field tunability of the transmon qubit allows us to perform power calibration in the entire operational  range of the grAlPA, as shown in \cref{Attn_SionLudi}. The frequency dependence of the attenuation is attributed to the frequency-dependent insertion loss of the coaxial cables. We observe an overall decrease in attenuation of $\approx$~5~dB from room temperature to $30$~mK, which can be explained by the change of resistivity in the coaxial cables. 

\section{Comparison of the reflection amplitude with and without pump}
\label{A_Pump_Off}

For our definition of gain, we compare the reflection amplitude $\mathrm{S_{11}}$ of each hybridized mode while having the grAlPA pump on and off. Residual internal losses produce a dip in $\mathrm{S_{11}}$ close to the resonance frequency in the undriven case, giving an apparent assymetry in the gain profiles. This effect is more evident at $B_\mathrm{||}$~=~1~T because of the proximity of the gain profiles to the initial resonance of the dimer modes (see Fig. 3(b) in the main text). In \cref{Gain_pumo_off} we show a comparison of the raw $\mathrm{S_{11}}$ measured in both the driven and undriven case at $B_\mathrm{||}$~=~1~T. We observe that when the grAlPA pump is on the profile of each mode resembles the expected Lorentzian shape. 

\section{Measurement setup}
\label{A_SetUp}
We use three different experimental setups, depicted in \cref{SetUp}. For the measurements presented in the main text and in Appendices \ref{A_LinCharac}-\ref{A_Magnetometer}, we use the configuration shown in \cref{SetUp}(a). The setup is adapted to perform both the input line-attenuation calibration via resonant fluorescence of a transmon qubit (see Appendix \ref{A_Magnetometer}) and the characterization of the grAlPA, using the same sample holder (see Fig. 1(a)-(b) in the main text) and microwave connections. The response of each device under test (DUT~=~qubit or grAlPA) is characterized in separate cooldowns using single-port reflection measurements. An input signal generated by a Vector Network Analyzer (VNA) is attenuated by a series of two nominally 30~dB attenuators thermalized to the 4~K and 30~mK plates of our dilution refrigerator, respectively. The remaining microwave components and connections in the input line provide an additional attenuation of about 30~dB. The output signal of the DUT is routed by a cryogenic circulator to a 2-stage isolator and a 12~GHz low-pass filter at the mixing chamber, and then amplified by a HEMT amplifier thermalized at 4~K \cite{HEMT_LNF} and a commercial room temperature amplifier. We connect the isolator and the HEMT with a NbTi superconducting cable with negligible insertion loss. We couple the pump tone using a directional coupler at room temperature. The output signal is split by a Wilkinson power divider and directed to the receiving port of the VNA and the input of a spectrum Analyzer (SA). The setup is equipped with a 2D vector magnet thermalized at 4~K, designed to produce in-plane magnetic fields up to 1~T and out-of-plane magnetic fields of the order of 100~mT.

In the second setup configuration, shown in \cref{SetUp}(b), the transmon qubit and the grAlPA are connected and measured in the same cooldown. They are mounted in separate sample holders and placed in different magnetic shields. The pump tone is now coupled at the lowest temperature stage of the dilution refrigerator using a cryogenic directional coupler. We estimate an insertion loss about 1.5~dB between the output of the transmon and the input of the grAlPA. This comes from the circulators (0.4~dB measured at 77~K by the supplier~\cite{Circulator_LNF}), directional coupler (0.5~dB measured at 10~mK~\cite{Dir_Coupler}) and microwave connections (30~cm with $\approx$~2~dB/m measured at 300~K~\cite{Microwave_Cables}) used in the setup. In this case only the qubit is magnetic field biased using a superconducting coil attached to the sample holder.

We use the setup in \cref{SetUp}(c) to extract the quantum efficiency of the grAlPA from qubit measurements (see Fig. 4(c)-(e) in the main text).  Compared to \cref{SetUp}(b), the GFQ is mounted inside a homemade Al magnetic shielding, there is an additional isolator (with 0.2~dB insertion loss~\cite{Circulator_LNF}) and microwave connections (with $\approx$~30~cm total length) between the grAlPA and the GFQ in the mixing chamber. To perform time domain measurements we use an Octave$^{\text{\textregistered}}$
 up/down conversion unit and a OPX+$^{\text{\textregistered}}$ FPGA board \cite{Octave}, connected at room temperature with 30~dB attenuators at the Octave$^{\text{\textregistered}}$ input and output ports.  

\section{Dependence of SNR improvement with pump power and frequency}
\label{A_SNR_3D}

In \cref{SNR_3D} we present measured grAlPA $\Delta$SNR as a function of pump power $P\mathrm{_p}$ and signal-pump detuning $\Delta f\mathrm{_p}$. The data was obtained employing the setup shown in \cref{SetUp}(b). We define the operational region of the amplifier as the set of pump powers and frequencies that provide stable $\Delta$SNR $\geq$ 10~dB.

\section{Noise performance in a secondary experimental setup using a transmon qubit}
\label{A_Noise_BF}

\begin{figure}[!t]
\begin{center}
\includegraphics[width = 1\columnwidth]{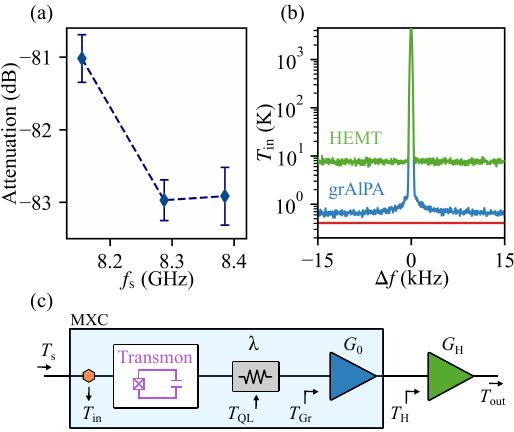}
\caption{\textbf{Noise performance of the grAlPA in the secondary measurement setup.} \textbf{(a)}~Input power calibration of the measurement setup shown in \cref{SetUp}(b). The attenuation is calculated using the fluorescence of the transmon qubit explained in Appendix \ref{A_Magnetometer}. \textbf{(b)}~Input-referred noise temperature as a function of the detuning $\Delta f$ from a power-calibrated tone at $f\mathrm{_s}$~=~8.469~GHz. The green and blue solid lines represent measurements with the grAlPA pump off and on, respectively. We attribute the increase in input-refereed noise compared to Fig. 4(a) of the main text to the insertion loss $\lambda$ between the qubit and the grAlPA. \textbf{(c)}~The losses between the qubit and the grAlPA are modeled by an effective attenuator thermalized to the mixing chamber of the dilution refrigerator. The orange marker indicates the point up to which we calibrate the input power.}
\label{Noise_BF_plot}
\end{center}
\end{figure}

In the following, we explain the noise characterization of the grAlPA performed in the setup illustrated in \cref{SetUp}(b). In this case the transmon qubit used for the input power calibration (see Appendix \ref{A_Magnetometer}) is connected before the amplifier in the same cooldown. Figure~\ref{Noise_BF_plot}(a) shows the attenuation calculated within the operational range of our amplifier in the absence of magnetic field. Figure~\ref{Noise_BF_plot}(b) shows the input referred noise temperature measured using a pilot tone at $f\mathrm{_s}$~=~8.469~GHz. When the amplifier is tuned to produce a maximum gain of $G\mathrm{_0}$ = 20~dB, we achieve $\Delta$SNR~$\approx$~10~dB and a noise performance approaching the standard quantum limit, in agreement with the results shown in Fig. 4(b) of the main text. The increase of the input-referred noise in this experiment, is attributed to the additional insertion loss between the transmon qubit and the grAlPA. 

\begin{figure}[!t]
\begin{center}
\includegraphics[width = 1\columnwidth]{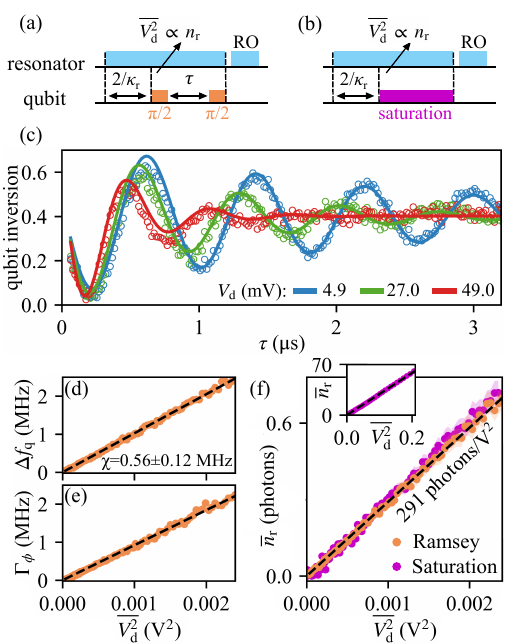}
\caption{\textbf{AC Stark shift, measurement induced dephasing and power calibration using a GFQ.} \textbf{(a)}-\textbf{(b)}~Pulse sequence used to measured the AC stark shift of the GFQ using Ramsey interferometry and Two-tone spectroscopy respectively. To reach steady-state, the resonator is populated $2/\kappa\mathrm{_r}$~=~260~ns before the start of the qubit manipulation. The readout pulse (RO) is the same as in Fig. 4(c)-(e) of the main text. \textbf{(c)} Example of Ramsey fringes obtained using the pulse sequence shown in panel (a) for three different readout voltages $V\mathrm{_{d}}$. The solid lines are fits using a sinusoidal function with an exponential decaying envelope. \textbf{(d)}-\textbf{(e)} AC stark shift $\Delta f\mathrm{_q}$ and dephasing rate $\Gamma\mathrm{_\phi}$ as a function of $\overline{V\mathrm{^2_{d}}}$ obtained from Ramsey experiments. The black dashed lines represent fits of Eqs. (\ref{eq_AC_stark})-(\ref{eq_Dephasing_rate}), from which we extract \corrs{$\chi$~=~0.56~$\pm$~0.12~MHz and $c$~=~290~$\pm$~35~photons/$\mathrm{V^2}$}. \textbf{(d)} Resonator occupation number $\overline{n}_\mathrm{_r}$ as a function of $\overline{V\mathrm{^2_{d}}}$, calculated from Ramsey interferometry (orange) and two-tone spectroscopy (purple). The black dashed line comes from the fits shown in panels (d) and (e). The inset shows that the power calibration remains valid up to $\overline{n}_\mathrm{_r} \approx$~70.}
\label{AC_stark_calibration}
\end{center}
\end{figure}

To account for the additional losses between the qubit and the amplifier, we assume that all added components can be modelled as an effective attenuator thermalized to the mixing chamber of the dilution refrigerator ($T\mathrm{_{Th}} \approx$~15~mK). A schematics is presented in \cref{Noise_BF_plot}(c). Using a beam-splitter approximation for the effective attenuator, we derive the relation between the noise referred back to the input of the transmon qubit $T\mathrm{_{in}}$ and the noise added by grAlPA $T\mathrm{_{gr}}$ as

\begin{equation}
T\mathrm{_{in}} = \frac{T\mathrm{_{out}}}{\lambda G\mathrm{_0} G\mathrm{_H}}  = T\mathrm{_{s}} + \left(\frac{1-\lambda}{\lambda}\right)T\mathrm{_{QL}} +\frac{T\mathrm{_{gr}}}{\lambda}+\frac{T\mathrm{_{H}}}{\lambda G\mathrm{_0}},
\label{eq_effective_insertion_Loss}
\end{equation}

where $\lambda$ is the insertion loss of the effective attenuator, $T\mathrm{_{QL}}$ corresponds to vacuum fluctuations, $T\mathrm{_{s}}$ the noise accompanying the pilot tone and $G\mathrm{_{H}}$, $T\mathrm{_{H}}$ the gain and added noise of the HEMT, respectively. Assuming that $\lambda$~=~-1.5~dB is the nominal insertion loss of the microwave components computed in Appendix \ref{A_SetUp}, $T\mathrm{_{s}}$~=~$T\mathrm{_{QL}}$ and taking $T\mathrm{_{H}}$ from \cref{Noise_BF_plot}(b), we obtain $T\mathrm{_{gr}}$~=~214~$\pm$~160~~mK. The error comes from the power calibration and $\pm$~0.5~dB uncertainty in the estimation of $\lambda$. Converting $T\mathrm{_{gr}}$ into photon number scale, we determine that the grAlPA adds $n\mathrm{_{gr}}$~=~0.53~$\pm$~0.40 noise photons to the measurement chain, consistent with the results shown in Fig 4(b) of the main text.

\section{Power calibration using the AC stark shift of a Generalized Flux Qubit}
\label{A_AC_stark}

\begin{table*}[t]
\corrs{
\caption{Comparison of grAlPA noise performance for three different power calibrations: (I) fluorescence of a transmon qubit measured in separate cooldowns, (II) fluorescence of a transmon qubit measured in the same cooldown and (III) AC stark shift of a GFQ. The uncertainty of the quantum efficiency is upper bounded to 100$\%$.}}
\begin{center}
\begin{tabular}{  p{1.5cm} | p{2.5cm}  p{2.5cm}  p{2cm}  p{2cm}  } 
\hline 
 &  \multicolumn{4}{ c}{ \corrs{Power calibration method}} \Tstruttop\Bstruttop  \\  \hline \\[-0.25cm]
\hfil  & \hfil (I) at $B_\mathrm{||}$ = 0~T & \hfil  (I) at $B_\mathrm{||}$ = 1~T & \hfil (II) & \hfil (III) \Tstrut\Bstrut \\[0.1cm]
\hline \\[-0.25cm]
\hfil $\eta$ ($\%$) & \hfil 76$\mathrm{^{+24}_{-36}}$ &\hfil 57 $\pm$ 27 & \hfil 62 $\pm$ 14 &\hfil 43 $\pm$ 9 \Tstrut\Bstrut \\[0.1cm] 
\hfil $\eta\mathrm{_{gr}}$ ($\%$) & \hfil 70$\mathrm{^{+30}_{-62}}$  &\hfil 46 $\pm$ 35 & \hfil 94$\mathrm{^{+6}_{-71}}$ &\hfil 92$\mathrm{^{+8}_{-22}}$ \Tstrut\Bstrut \\[0.1cm] 
\hfil $n\mathrm{_{gr}}$ & \hfil 0.71 $\pm$ 0.63 &\hfil 1.08 $\pm$ 0.83 & \hfil 0.53 $\pm$ 0.40 &\hfil 0.54 $\pm$ 0.13 \Tstrut\Bstrut \\ [0.1cm]
\hline
\end{tabular}
\end{center}
\label{tab:Comparison_noise_performance}
\end{table*}

We obtain the calibration of the photon number circulating in the readout resonator by measuring the AC Stark shift of the GFQ using two different experiments, as depicted in \cref{AC_stark_calibration}(a) and (b). For the first one, we perform Ramsey interferometry measurements while driving the readout resonator with a room temperature voltage amplitude $V\mathrm{_{d}}$. The driving tone populates the resonator with an average photon number $\overline{n}_\mathrm{_r}$, resulting in a shift in the qubit frequency
\begin{equation}
\Delta f_{q} = \chi \overline{n}_\mathrm{_r} = \chi \left(c \overline{V\mathrm{^2_{d}}}\right)
\label{eq_AC_stark}
\end{equation}
and an increase in the dephasing rate \cite{Gambetta2006Oct}
\begin{equation}
\Delta \Gamma_{\phi} = \frac{2 \chi^2}{\kappa} \overline{n}_\mathrm{_r},
\label{eq_Dephasing_rate}
\end{equation}
where $\kappa/2\pi$~=~1.25~MHz and $\chi$ are the linewidth and dispersive shift of the readout resonator respectively, and $c$ is the proportionality constant constant relating $\overline{n}_\mathrm{_r}$ and $\overline{V\mathrm{^2_{d}}}$. The results of the Ramsey experiments for three different values of $V\mathrm{_{d}}$ are shown in \cref{AC_stark_calibration}(b), where the solid lines represent fits using a sinusoidal function with an exponential decaying envelope, from which we extract $\Delta f_{q}$ and $\Delta \Gamma_{\phi}$. In \cref{AC_stark_calibration}(d)-(e) we plot the extracted values of $\Delta f_{q}$ and $\Delta \Gamma_{\phi}$ as a function of $\overline{V\mathrm{^2_{d}}}$. We fit both quantities simultaneously using Eqs. (\ref{eq_AC_stark})-(\ref{eq_Dephasing_rate}) and obtain \corrs{$\chi$~=~0.56~$\pm$~0.12~MHz and $c$~=~290~$\pm$~35~photons/$\mathrm{V^2}$}. Note that the calibration using Ramsey experiments is limited to photon numbers below $\overline{n}_\mathrm{_r} \approx$ 1, as the qubit $T\mathrm{_2}$ time decreases below the resolution time of the readout electronics at higher readout powers. To overcome this limitation we use two-tone spectroscopy with a qubit saturation pulse while populating the resonator with $\overline{n}_\mathrm{_r}$ photons, as illustrated in \cref{AC_stark_calibration}(b). As shown in \cref{AC_stark_calibration}(f), for low readout powers, the two calibration methods agree. Moreover, the inset of \cref{AC_stark_calibration}(f) shows that the calibration remains valid for $\overline{n}_\mathrm{_r} \approx$~70.

\section{Temperature of the GFQ}
\label{A_qubit_Temperature}

We use the IQ histogram of the GFQ in equilibrium (see Fig. 4(c) in the main text) to extract an estimation of the effective qubit temperature using the formula,
\begin{equation}
T\mathrm{_q}=\frac{2\pi \hbar f\mathrm{_q}}{k\mathrm{_B}\mathrm{ln}\left(N\mathrm{_g}/N\mathrm{_e}\right)},
\label{eq_qubit_temperature}
\end{equation}
where $N_\mathrm{g,e}$ are qubit ground and first excited state populations and $f_\mathrm{_q}$~=~6.0195~GHz its frequency. We neglect the contributions of higher-energy states as they account less than 0.01$\%$ of the qubit excitations. We obtain an effective qubit temperature $T\mathrm{_q}$~=~56~mK. 

\section{Quantum efficiency budget}
\label{A_quantum_efficiency}

The readout chain quantum efficiency obtained in the main text using the GFQ is \corrs{$\eta$~=~43~$\pm$~9$\%$}. We identify four different contributions: $\eta$=$\eta\mathrm{_{r}}\cdot\,\eta\mathrm{_{IL}}\cdot\,\eta\mathrm{_{H}}\cdot\,\eta\mathrm{_{gr}}$, where $\eta\mathrm{_{r}}$ is the internal loss of the GFQ readout resonator, $\eta\mathrm{_{IL}}$ is the insertion loss between the GFQ and the grAlPA, $\eta\mathrm{_{H}}$ is the HEMT (input referred) noise scaled by the grAlPA gain and $\eta\mathrm{_{gr}}$ is the intrisic quantum efficiency of the grAlPA.

In the ideal case, the noise at the output of the GFQ readout resonator is given by vacuum fluctuations $T_\mathrm{QL}$. Coupling to unwanted loss channels translates into an additional noise factor $T_\mathrm{r}$~=~$T_\mathrm{QL}\left(\gamma\mathrm{_r}/\kappa\mathrm{_r}\right)$ \cite{IvanThesis2022}, where $\gamma\mathrm{_r}/2\pi$~=~0.17~MHz, $\kappa\mathrm{_r}/2\pi$~=~1.25~MHz are the resonator internal and external loss rates, respectively. In our case, the quantum efficiency at the output of the resonator is $\eta\mathrm{_{r}}$~=~$T_\mathrm{QL}/\left(T_\mathrm{QL} + T_\mathrm{r} \right)$~=~0.88.

To calculate $\eta\mathrm{_{IL}}$, we use an effective attenuator model, as sketched in \cref{Noise_BF_plot}(c) in Appendix \ref{A_Noise_BF}. The non-zero insertion loss $\lambda$ increases the output noise to $T_\mathrm{IL}$~=~$\lambda T_\mathrm{QL} + (1-\lambda) T_\mathrm{QL}$. We use the nominal attenuation $\lambda$~=~-2.3~dB (see Appendix \ref{A_SetUp}) to compute $\eta\mathrm{_{IL}}$~=~$\lambda T_\mathrm{QL}/T_\mathrm{IL}$~=~0.59~$\pm$~0.04. The error corresponds to $\pm$~0.5~dB uncertainty in the insertion loss.

The noise of the HEMT referred to the input of the readout chain is given by $T_\mathrm{H}/\lambda\cdot G\mathrm{_0}$. In the ideal case when this contribution is negligible, the GFQ input referred noise $2T_\mathrm{QL}$ is given by the sum of vacuum fluctuations and the grAlPA quantum-limited noise. Using $T_\mathrm{H}$ from the noise floor of the pump-off PSD in \cref{Noise_BF_plot}(b), we extract $\eta\mathrm{_{H}}$~=~$2T_\mathrm{QL}/\left(2T_\mathrm{QL} + T_\mathrm{H}/\lambda\cdot G\mathrm{_0} \right)$~=~0.91.

Combining all contributions we obtain an intrinsic grAlPA quantum efficiency \corrs{$\eta\mathrm{_{gr}}$~=~92$\mathrm{^{+8}_{-22}}\%$}. This corresponds to \corrs{$n\mathrm{_{gr}}$~=~$n\mathrm{_{QL}}$/$\eta\mathrm{_{gr}}$~=~0.54~$\pm$~0.13 added noise photons}, close to the quantum limit $n\mathrm{_{QL}}$~=~0.5. Part of the $\eta\mathrm{_{gr}}$ reduction can be explained by internal losses of the dimer modes. A fit of the grAlPA modes performed in the same cooldown of the data shown in Fig. 4(c)-(e) of the main text, gives $\kappa\mathrm{_-}$~=~19.4~$\pm$~4.1~MHz, $\gamma\mathrm{_-}$~=~0.3$\mathrm{^{+4.1}_{-0.3}}$~MHz, $\kappa\mathrm{_+}$~=~25.0$\mathrm{^{+4.9}_{-5.9}}$~MHz and $\gamma\mathrm{_+}$~=~2.0$\mathrm{^{+5.3}_{-2.0}}$~MHz, where the errors where calculated from the Fano uncertainty of the measurements \cite{Rieger2023fano}. They contribute to an additional noise factor $T_\mathrm{\gamma_{\pm}}$~=~$T_\mathrm{QL}\left(\gamma\mathrm{_+}/\kappa\mathrm{_+}\right)$~+~$T_\mathrm{QL}\left(\gamma\mathrm{_-}/\kappa\mathrm{_-}\right)$, from which we extract $\eta\mathrm{_\mathrm{\gamma_{\pm}}}$~=~$T_\mathrm{QL}/\left(T_\mathrm{QL} + T_\mathrm{\gamma_{\pm}} \right)$~=~0.91$\mathrm{^{+0.09}_{-0.25}}$. Therefore internal losses might fully account for the reduction in the grAlPA intrinsic quantum efficiency $\eta\mathrm{_{gr}}$.

\section{Comparisson of noise performance using different power calibrations}
\label{A_quantum_efficiency_summary}

In Table \ref{tab:Comparison_noise_performance} we show a comparison of the total measurement efficiency $\eta$, intrinsic grAlPA quentum efficiency $\eta\mathrm{_{gr}}$ and amount of added noise $n\mathrm{_{gr}}$ for the three different power calibrations used in this work: (I) fluorescence of a transmon qubit measured in separate cooldowns (see main text and Appendix \ref{A_Magnetometer}), (II) fluoresncence of a transmon qubit measured in the same cooldown (see Appendix \ref{A_Noise_BF}) and (III) AC stark shift of a GFQ (see Appendix \ref{A_AC_stark}). The values of $\eta$ for method (I) and (II) are calculated at the plane of the power calibration. We compute $\eta\mathrm{_{gr}}$ and $n\mathrm{_{gr}}$ for method (I) from the data presented in Fig.~4(b) of the main text and using the relations $n\mathrm{_{gr}}$~=~$n\mathrm{_{in}}$-$n\mathrm{_{QL}}$-$n\mathrm{_{H}}/G\mathrm{_0}$ and $\eta\mathrm{_{gr}}$~=~$n\mathrm{_{QL}}/n\mathrm{_{gr}}$ respectively. We calculate $n\mathrm{_{H}}$ with the formula $n\mathrm{_{H}}=k\mathrm{_B}\overline{T}\mathrm{_{H}}$/$hf\mathrm{_s}$, where $\overline{T}\mathrm{_{H}}$ is the noise temperature floor extracted from the pump-off PSD in Fig.~4(a) of the main text and $f\mathrm{_s}$ the frequency of the power-calibrated pilot tone. We observe comparable noise performance for all methods, which validates the near quantum-limited efficiency of the grAlPA.

\bibliography{GrAlPA}

\end{document}